\documentclass[10pt,twocolumn]{article}
\usepackage{amsmath,amssymb,amsthm}
\usepackage[hidelinks]{hyperref}
\hypersetup{
  pdftitle={Beyond Proof of Possession: A Relational Theory of Authority},
  pdfauthor={Nicola Gallo}
}
\usepackage{microtype}
\usepackage{booktabs}
\usepackage{array}
\usepackage[letterpaper,margin=0.72in]{geometry}
\newcommand{\sem}[1]{\left[\!\left[#1\right]\!\right]}
\newcommand{\Rel}{\mathcal{R}}
\newcommand{\Profiles}{\mathcal{Q}}
\newcommand{\Transitions}{\mathcal{T}}
\newcommand{\Lineages}{\mathcal{L}}
\newcommand{\Pred}{\mathsf{Pred}}
\newcommand{\exec}{\operatorname{exec}}
\newcommand{\ValidR}{\operatorname{Valid}_{R}}
\newcommand{\PoR}{\operatorname{PoR}}
\newcommand{\PoC}{\operatorname{PoC}}
\newcommand{\PoP}{\operatorname{PoP}}

\title{Beyond Proof of Possession:\\
A Relational Theory of Authority}
\author{Nicola Gallo}
\date{25 September 2026}

\newtheorem{definition}{Definition}
\newtheorem{lemma}{Lemma}
\newtheorem{theorem}{Theorem}
\newtheorem{proposition}{Proposition}
\newtheorem{corollary}{Corollary}
\newtheorem{remark}{Remark}
\newtheorem{assumption}{Assumption}

\begin{document}
\maketitle

\begin{abstract}
Proof-of-Possession (PoP) establishes evidence that an executor controls an
artifact such as a key, token, credential, or capability. It may support holder
binding, authentication, profile evidence, or a concrete continuation
protocol, but control of an artifact alone determines neither the executor's
relational profile nor the causal history of an execution occurrence. We
separate PoP from Proof of Relationship (PoR), the semantic property that one
execution occurrence validly continues another, and from authorization.
Authorization is occurrence-sensitive and indexed by the occurrence's causal
history: a transition is evaluated from PoR, non-expansion of its authority
context, satisfaction and refinement of its relational envelope, and soundly
established executor profiles.
The foundational change is a change in the semantic domain of authorization:
from holders and authority-bearing artifacts alone to execution occurrences
situated in causal histories.
We make that distinction comparable across mechanism families by giving an
authorization model an observation-based occurrence verdict and a separate
transition-admissibility relation. Kernel inclusion defines a partial order on
observation-kernel classes through which concrete realizations can be compared,
while transition invariants form a second axis; neither is a ranking of
security.

Building on Provenance Identity Continuity (PIC) and Proof-of-Continuity (PoC),
we distinguish execution steps from the executors that realize them and
classify executors by authorization-relevant equivalence relations. Their
quotients form relational profile spaces, while positive predicates over
profile transitions form a distributive lattice ordered by implication. This
adds relational attenuation to authority attenuation: along one valid causal
history, \(C_{i+1}\subseteq C_i\) and
\(\rho_{i+1}\Rightarrow\rho_i\). Causal history grows by prefix extension
while authority and relational freedom shrink. We prove dual monotonicity,
combined closure, and origin relational closure.

Because authorization is defined over occurrences situated in causal
histories, two occurrences of the same authenticated executor, with the same
profile and possession evidence, need not be authorization-equivalent. Within
a valid PIC lineage, independent authority held by a deputy cannot be imported
as a valid exercise outside the origin context. Excluding request-caused
effects presented outside that lineage additionally requires complete
mediation and causal attribution.

Finally, we prove that the executor-dependent decision factors through the
relational profile and is invariant under sound mechanisms that establish the
same profile. Successor-side PoP is therefore a possible evidence mechanism,
not a necessary semantic premise of authorization. Conditional on explicit
Witness Assumptions, a boundary-witnessed checkpoint construction realizes the
premises without successor-side PoP. Boundary signatures provide checkpoint
evidence, profile evidence may rely on additional trust roots, and the
construction does not independently establish its environmental assumptions.
\end{abstract}

\section{Introduction}

Many deployed authorization mechanisms attach authority to something that an
actor possesses or controls: a bearer token, a sender-constrained token, a key,
a credential, a capability, or an ambient privilege. Such mechanisms are often
effective local authorization tools. Their common abstraction, however, is
spatial: the decision is made from authority that is available to the actor at
the point of use.

Proof-of-Continuity~\cite{gallo2026poc} introduced a different axis. A later execution step is valid
only if it is causally related to its predecessor and carries no more authority
than the predecessor received. Formally, for a causal chain
\[
s_0 \rightarrow s_1 \rightarrow \cdots \rightarrow s_n,
\]
each valid hop satisfies
\[
\PoR(s_i,s_{i+1})
\qquad\text{and}\qquad
C_{i+1}\subseteq C_i,
\]
where \(C_i\subseteq O\times R\) is the authority context carried at step
\(s_i\)~\cite{gallo2026poc}. This gives
\[
C_n\subseteq C_{n-1}\subseteq\cdots\subseteq C_0
\]
and therefore origin-bounded authority: a privilege absent from \(C_0\) cannot
appear later in the same valid lineage.

That result leaves a further question open. If PoR is explicitly a
\emph{relationship} between execution occurrences, what mathematical structure
does that relationship carry? More specifically, can authorization be expressed
over relational characteristics of the execution rather than over possession by
a particular holder?

The foundational change is not a new credential format but a change in the
domain over which authorization is defined: from a decision described only by
holders and authority-bearing artifacts to one evaluated over execution
occurrences situated in causal histories. Two occurrences of the same
authenticated executor, presenting the same possession evidence, need not be
authorization-equivalent when they occur in different histories. The same
agent may present the same key, token and profile in two requests; a write may
be admissible in the history whose origin context contains it and inadmissible
in the history whose origin context does not. The difference is not the holder
or artifact but the causal history of the occurrence.

Thus possessing authority is not the same as being the execution occurrence
permitted to exercise it in the current causal history. Bare possession
determines neither an occurrence's causal lineage nor
\(\PoR(s_i,s_{i+1})\).
Authority is exercisable not merely because an executor possesses it, but
because the current occurrence is an admitted continuation of a history in
which that authority remains valid.

To answer the question above, this paper introduces a relational space for
executors. Let \(X\) be the set of executor instances. An
authorization-relevant equivalence relation \(\sim_j\) partitions \(X\) into
classes of executors that are indistinguishable with respect to one relational
characteristic. The quotient
\[
\Rel_j := X/{\sim_j}
\]
therefore forgets all executor differences irrelevant to that characteristic.
Different relations induce different quotient views, so the same executor may
belong to different equivalence classes under different relations.

We then define relational predicates over combinations of these quotient views.
Predicates compose with logical operators such as \(\land\) and \(\lor\), and
logical implication induces an attenuation order:
\[
\rho' \preceq_R \rho
\quad\Longleftrightarrow\quad
\rho' \Rightarrow \rho.
\]
Thus a successor may preserve or restrict a relational authorization envelope,
but may not expand it.

The result is a dual monotonicity. Along one causal chain,
\[
C_{i+1}\subseteq C_i
\]
restricts authority, while
\[
\rho_{i+1}\Rightarrow\rho_i
\]
restricts relational admissibility. The lineage prefix itself moves in the
opposite direction: causal history accumulates. Hence continuity has a simple
structural form:
\[
\begin{gathered}
\text{history grows, while authority}\\
\text{and relational freedom shrink.}
\end{gathered}
\]

The paper makes seven contributions:
\begin{enumerate}
    \item it separates execution steps from executors and gives executor
    relations a quotient-space semantics;
    \item it defines multiple relational views of the same executor and a
    combined relational profile space;
    \item it defines a positive relational predicate lattice with refinement
    ordered by implication;
    \item it proves dual monotonicity and relational closure along a PoR-linked
    causal chain;
    \item it shows that the lineage-forgetting projection of the previous
    paper is itself a relational quotient, so that its lineage-invariance
    result for possession-based policies is an instance of the factorization property
    developed here;
    \item it places artifact-centered, sender-constrained, attenuating
    delegation and PIC-compatible authorization in a common occurrence domain,
    defines a refinement order on observation-kernel classes through which
    such mechanisms can be compared, and separately classifies their
    transition invariants; and
    \item it proves that the executor-dependent component of authorization
    factors through the relational profile, formalizes invariance under sound
    profile-evidence mechanisms that produce the same profile, and gives,
    relative to explicit Witness Assumptions, a construction in which neither
    profile nor PoR evidence requires successor-side possession.
\end{enumerate}

Monotone attenuation along a chain is not itself new; capability tokens and
delegation-chain systems already provide it (Section~\ref{sec:related}). Nor
do the mechanism families considered here form a total order. The contribution
is a common mathematical comparison that separates what a mechanism observes
from the invariants it enforces, together with monotone attenuation of an
\emph{executor-relational admissibility space} composed with authority
attenuation and causal continuity. Section~\ref{sec:picx} maps a restricted
subset of the premises onto an open-source implementation and states the gaps
explicitly.

The result does not claim that PoP is useless, nor that deployed
possession-based authorization systems do not authorize. Rather, it separates
two roles that are often collapsed: \emph{evidence that an actor controls
something} and \emph{the semantic condition that makes a continuation
authorized}. In the model developed here, possession may help establish the
former. Authentication fixes executor/profile facts; PoR establishes causal
continuation; and \(C_i\) together with \(\rho_i\) bounds the authorization
envelope.

As an intuition rather than a definition, PoP answers what artifact or key an
entity demonstrates control over; PoR answers what execution occurrence is
being continued; and authorization answers what that occurrence may exercise
here, in its causal history. PoR is not by itself the authorization decision,
and a PoP transcript is not by itself a relational profile or a causal history.

The picture is one of two layers. At admission, a sound evidence mechanism
establishes facts about the executor and fixes its profile \(q_J(x)\). PoP may
provide holder-binding evidence only when combined with a trusted binding from
the possessed artifact to the executor and the asserted classes; the evidence
mechanism need not use successor-side PoP. Once the profile is established,
PoR determines whether the occurrence is a causal continuation, \(C_i\) bounds
which authority may continue, and \(\rho_i\) both tests and bounds the executor
relationships through which it may continue.
Capability constructions may designate, carry and attenuate authority; PoP
mechanisms may establish control of authority-bearing material. PIC adds the
separate semantic question whether this occurrence is a valid causal
continuation under non-expanding authority and relational state. These
dimensions may be combined, and the carrier does not determine the semantic
class.
Treating possession as the ground of authority collapses the two layers. A
single-hop mechanism may deliberately identify designation and authority.
Across hops, however, any decision that factors through a lineage-forgetting
observation cannot distinguish occurrences that differ only in causal history
(Remark~\ref{rem:instance}).

\section{Background and Previous Result}

\subsection{Authority as a causal continuation}

Let \(O\) be a set of operations and \(R\) a set of resources. A privilege is
\[
(o,r)\in O\times R.
\]
A discrete execution is a causal chain
\[
\pi=\langle (s_0,C_0),(s_1,C_1),\ldots,(s_n,C_n)\rangle,
\]
where \(s_i\) is an execution step and
\[
C_i\subseteq O\times R
\]
is its authority context.

The previous PoC model defines a single-hop Proof of Relationship by
\[
\PoR(s_i,s_{i+1}),
\]
which holds when \(s_{i+1}\) is a valid causal continuation of \(s_i\) within
the same execution lineage~\cite{gallo2026poc}. A transition is valid when it
is both causally related and non-expansive:
\[
\operatorname{Valid}(s_i,C_i,s_{i+1},C_{i+1})
\Longleftrightarrow
\PoR(s_i,s_{i+1})
\land
C_{i+1}\subseteq C_i.
\]

The whole chain satisfies Proof-of-Continuity when every adjacent transition is
valid:
\[
\PoC(\pi)
\Longleftrightarrow
\bigwedge_{i=0}^{n-1}
\left(
\PoR(s_i,s_{i+1})
\land
C_{i+1}\subseteq C_i
\right).
\]

\begin{proposition}[Origin-bounded authority, previous result]
If \(\PoC(\pi)\) holds, then
\[
C_n\subseteq C_{n-1}\subseteq\cdots\subseteq C_0.
\]
Consequently, if \((o,r)\notin C_0\), then
\[
(o,r)\notin C_i
\]
for every \(i\).
\end{proposition}

\begin{proof}
Immediate by transitivity of set inclusion.
\end{proof}

\subsection{Confused deputy as authority/causality mismatch}

The classical confused deputy occurs when a request causes a deputy to exercise
authority that does not belong to the request's authority context. Capabilities
address the classical single-invocation form by fusing designation and
authority~\cite{hardy1988confused,miller2003capability}. In a distributed or
multi-step execution, however, an executor may hold several unrelated authority
sources simultaneously. The PIC project's N+1 Invalid State problem statement
documents this authority-mixing motivation~\cite{picAuthorityMixing}.

Let a request originate with authority context \(C_0\). If an executor later
uses a privilege
\[
(o,r)\notin C_0
\]
because it independently possesses that privilege, then authority and causality
have diverged. The previous PoC result rules this out as valid behavior because
every valid descendant context is a subset of \(C_0\).

The present work preserves this result and adds a second closure dimension: a
valid descendant may not expand the relational class of executions through
which origin authority is permitted to continue.

\section{Execution Steps and Executors}

The previous model uses \(s_i\) for a formal execution step. Operationally,
that step is realized by an executor such as a service, workload, process,
function, tool, agent, or hardware component.

\begin{definition}[Executor space]
Let
\[
S
\]
be the set of execution steps and let
\[
X
\]
be the set of executor instances. Define
\[
\exec:S\rightarrow X.
\]
For each \(s_i\in S\), write
\[
x_i:=\exec(s_i).
\]
\end{definition}

A causal chain
\[
s_0\rightarrow s_1\rightarrow\cdots\rightarrow s_n
\]
therefore induces an executor sequence
\[
x_0,x_1,\ldots,x_n.
\]

The distinction between step and executor is essential. The same executor may
realize several distinct occurrences:
\[
s_i\neq s_j
\qquad\text{while}\qquad
x_i=x_j.
\]
Conversely, a lineage may cross many different executors.

Authorization therefore cannot be reduced to executor identity alone. The
relevant object is an \emph{occurrence} of an executor in a causal execution and
the relationships under which that occurrence is admitted.

\section{Relational Equivalence and Executor Quotients}

\subsection{Equivalence relative to an authorization characteristic}

We now formalize the idea that two physically different executors may be
equivalent for one authorization purpose.

\begin{definition}[Authorization-relevant equivalence]
Let \(\sim_j\) be an equivalence relation on \(X\):
\[
\sim_j\;\subseteq X\times X.
\]
For \(x,y\in X\),
\[
x\sim_j y
\]
means that \(x\) and \(y\) are indistinguishable with respect to relational
characteristic \(j\).
\end{definition}

Since \(\sim_j\) is an equivalence relation, it is reflexive, symmetric, and
transitive. It partitions \(X\) into disjoint equivalence classes.

\begin{definition}[Equivalence class]
For \(x\in X\),
\[
[x]_{\sim_j}
:=
\{y\in X:y\sim_j x\}.
\]
\end{definition}

The class contains every executor that is indistinguishable from \(x\) under
the selected relation.

\begin{definition}[Relational quotient]
The quotient induced by \(\sim_j\) is
\[
\boxed{
\Rel_j := X/{\sim_j}
}
\]
with
\[
\Rel_j
=
\{
[x]_{\sim_j}:x\in X
\}.
\]
We call \(\Rel_j\) a \emph{relational quotient}.
\end{definition}

The quotient operation intentionally discards distinctions that the relation
declares irrelevant. It does \emph{not} assert that two concrete executors are
identical.

\subsection{A concrete quotient example}

Let
\[
X=\{a,b,c,d,e,f\}.
\]
Suppose \(\sim_1\) induces the partition
\[
\{a,b\},
\qquad
\{c,d,e\},
\qquad
\{f\}.
\]
Then
\[
\Rel_1
=
X/{\sim_1}
=
\big\{
\{a,b\},
\{c,d,e\},
\{f\}
\big\}.
\]

Since
\[
a\sim_1 b,
\]
we have
\[
[a]_{\sim_1}
=
[b]_{\sim_1}
=
\{a,b\}.
\]

Thus the quotient identifies \(a\) and \(b\) only with respect to the
characteristic represented by \(\sim_1\).

Now let a second relation \(\sim_2\) induce
\[
\{a,c,f\},
\qquad
\{b,d\},
\qquad
\{e\}.
\]
Then
\[
\Rel_2
=
X/{\sim_2}
=
\big\{
\{a,c,f\},
\{b,d\},
\{e\}
\big\}.
\]

The same executor \(a\) therefore has
\[
[a]_{\sim_1}
=
\{a,b\}
\]
but
\[
[a]_{\sim_2}
=
\{a,c,f\}.
\]

There is no contradiction. The same executor is being observed through two
different authorization-relevant relations.

\subsection{Quotient maps}

Each equivalence relation has a canonical quotient map
\[
q_j:X\rightarrow \Rel_j
\]
defined by
\[
q_j(x)=[x]_{\sim_j}.
\]

The map says: retain only the relational class of \(x\) and forget its
irrelevant concrete differences.

This gives the intended reading:
\[
\begin{gathered}
X/{\sim_j}
=\\
\text{``\(X\) viewed at the resolution}\\
\text{determined by \(\sim_j\)''.}
\end{gathered}
\]

\section{Multiple Relational Views}

A single executor may participate in several authorization-relevant
relationships. Let \(J\) index a family
\[
\{\sim_j\}_{j\in J}.
\]

Each relation produces its own quotient:
\[
\Rel_j=X/{\sim_j}.
\]

\begin{definition}[Relational profile map]
Define
\[
q_J:X\rightarrow \prod_{j\in J}\Rel_j
\]
by
\[
q_J(x)
=
\big(
[x]_{\sim_j}
\big)_{j\in J}.
\]
The realizable relational profile space is
\[
\Profiles_J
:=
q_J(X)
\subseteq
\prod_{j\in J}\Rel_j.
\]
\end{definition}

The use of the image \(q_J(X)\), rather than the full product, records only
combinations of quotient classes realized by some executor.

\subsection{Combined equivalence}

The family of relations also induces a combined equivalence:
\[
x\equiv_J y
\quad\Longleftrightarrow\quad
\forall j\in J,\;x\sim_j y.
\]
Equivalently,
\[
\equiv_J
=
\mathop{\cap}\limits_{j\in J}\sim_j.
\]

Two executors are therefore equivalent under \(J\) precisely when their entire
relational profiles agree:
\[
x\equiv_J y
\quad\Longleftrightarrow\quad
q_J(x)=q_J(y).
\]

\begin{proposition}[Factorization through a quotient]\label{prop:factorization}
Let \(Y\) be a set, \(\equiv\) an equivalence relation on \(Y\), and
\(q:Y\rightarrow Y/{\equiv}\) the canonical quotient map. A policy
\[
A:Y\rightarrow\{0,1\}
\]
is invariant under \(\equiv\), i.e.
\[
y\equiv y'
\Longrightarrow
A(y)=A(y'),
\]
if and only if there exists
\[
\bar A:Y/{\equiv}\rightarrow\{0,1\}
\]
such that
\[
A=\bar A\circ q.
\]
When it exists, \(\bar A\) is unique.
\end{proposition}

\begin{proof}
If \(A=\bar A\circ q\) and \(y\equiv y'\), then \(q(y)=q(y')\), hence
\(A(y)=A(y')\). Conversely, for a class \(c\in Y/{\equiv}\) choose any \(y\)
with \(q(y)=c\) and define \(\bar A(c):=A(y)\). Invariance makes this value
independent of the chosen representative, and \(A=\bar A\circ q\) by
construction. Uniqueness follows because \(q\) is surjective.
\end{proof}

For executors, take \(Y=X\) and \(\equiv\;=\;\equiv_J\). Since
\(x\equiv_J y\) iff \(q_J(x)=q_J(y)\), the profile map induces a bijection
\(X/{\equiv_J}\cong\Profiles_J\). A policy on executors is therefore invariant
under \(\equiv_J\) iff it factors as
\[
A=\bar A\circ q_J,
\qquad
\bar A:\Profiles_J\rightarrow\{0,1\}.
\]

This is the formal statement that an authorization rule may operate on
relational equivalence classes rather than on concrete executor identity. The
proposition is the standard universal property of a quotient; its interest here
lies in what it unifies.

\begin{remark}[The previous result as a quotient instance]\label{rem:instance}
In~\cite{gallo2026poc}, an event is an occurrence
\(e=(o,r,\ell)\in E:=O\times R\times\mathcal L_{\mathrm{prev}}\), where
\(\mathcal L_{\mathrm{prev}}\) is that paper's history space, and the projection
\(p:E\rightarrow O\times R\) forgets the lineage. Let \(\sim_{(o,r)}\) be the
kernel relation of \(p\),
\[
e\sim_{(o,r)}e'
\quad\Longleftrightarrow\quad
p(e)=p(e'),
\]
and \(q:E\rightarrow E/{\sim_{(o,r)}}\) its canonical quotient map. Since \(p\)
is surjective, there is a canonical bijection
\(\phi:E/{\sim_{(o,r)}}\rightarrow O\times R\) with \(p=\phi\circ q\). Up to
this bijection, \(p\) is the quotient map, and a policy factors through \(p\)
iff it factors through \(q\).
The possession-based policies defined there, which factor through \(p\) or
equivalently are lineage-invariant (Definition~7 of~\cite{gallo2026poc}), are
exactly the policies invariant under \(\sim_{(o,r)}\). The result that
lineage-invariant policies cannot individuate occurrences (Theorem~3
of~\cite{gallo2026poc}) is Proposition~\ref{prop:factorization} applied to
\(Y=E\) and \(\equiv\;=\;\sim_{(o,r)}\): two events in the same class receive the
same verdict. The companion result (Theorem~4 of~\cite{gallo2026poc})
establishes that a policy separating two events collapsed by the
lineage-forgetting projection must read finer, lineage-sensitive information.
This identifies the need for a finer observation; it does not by itself
establish the causal-continuity semantics developed here.

More generally, for any policy of the form \(P=\bar P\circ\operatorname{Obs}\),
\[
\operatorname{Obs}(s)=\operatorname{Obs}(s')
\Longrightarrow
P(s)=P(s').
\]
Thus a policy factoring through a lineage-forgetting observation cannot
distinguish two otherwise identical privilege uses solely by their different
causal histories. A capability construction that verifies an equivalent
lineage-sensitive continuation condition lies outside this invariant policy
class; the factorization result does not claim that every capability system is
lineage-invariant.

The present construction therefore does not add a structure beside that of the
previous paper; it contains it. The lineage-forgetting projection is one
relational quotient, and the executor relations \(\sim_j\) are further ones,
all governed by the same factorization property.
\end{remark}

\section{Relational Predicates}

The quotient spaces classify executors. Authorization additionally needs a
language for stating which relational profiles, and which transitions between
profiles, are admissible.

The two layers have different jobs. An equivalence relation \(\sim_j\) must be
reflexive, symmetric and transitive, while relations such as delegation,
acts-for, trust, parent and child, supervision, and causal precedence lack at
least one property required of an equivalence relation, although some may
satisfy one or more of them. The quotients are therefore
\emph{classification spaces}, not the
authorization relation itself. The authorization-relevant relation between a
predecessor and its successor is a predicate on
\(\Profiles_J\times\Profiles_J\), which may be directional and need not be an
equivalence. A supervisory relation, for instance, is an atom
\(\mathrm{Sup}(u,v)\) that holds when the class of \(u\) supervises the class
of \(v\).

\subsection{Transition profile space}

A causal hop involves two executors:
\[
x_i=\exec(s_i),
\qquad
x_{i+1}=\exec(s_{i+1}).
\]

Define the transition profile space
\[
\Transitions_J
:=
\Profiles_J\times\Profiles_J.
\]

The relational observation of a concrete hop is
\[
\tau_i
:=
\big(
q_J(x_i),
q_J(x_{i+1})
\big)
\in\Transitions_J.
\]

\subsection{Atomic and composite predicates}

\begin{definition}[Relational predicate]
A relational predicate is a function
\[
\rho:\Transitions_J\rightarrow\{0,1\}.
\]
Its semantic extension is
\[
\sem{\rho}
:=
\{\tau\in\Transitions_J:\rho(\tau)=1\}.
\]
\end{definition}

Let
\[
R_1,R_2,\ldots
\]
be atomic predicates over \(\Transitions_J\). We use the positive relational
language generated by
\[
\rho ::= R_k \mid (\rho\land\rho)\mid(\rho\lor\rho).
\]

We write \(\Pred(\Transitions_J)\) for the set of positive relational
predicates generated by this grammar from the atoms. Relational envelopes range
over \(\Pred(\Transitions_J)\), not over arbitrary functions
\(\Transitions_J\rightarrow\{0,1\}\).

\paragraph{Why no negation.}\label{sec:relpred}
The answer depends on what a boundary knows about an atom. Each atom is a fixed
subset of \(\Transitions_J\), and the profiles in \(\tau\) are verified. Some
atoms are decidable from \(\tau\) itself, such as equality or inequality of two
profile components. For those, the atom and its negation are equally
decidable, and nothing is lost by admitting \(\lnot\). The separation-of-duty
atom of Section~\ref{sec:sod} is of this kind, which is why it may contain
\(\notin\) and \(\neq\). For other atoms the extension is not known to the
boundary: a supervision relation that depends on an external organizational
chart, or a relation recorded in a delegation registry. The boundary then
learns membership only through evidence, and the truth of the atom must be kept
apart from the availability of evidence for it.

Let \(E\) be the evidence available to a boundary, and write
\(E\vdash R_k(\tau)\) when \(E\) supports \(R_k(\tau)\). The relation lifts to
positive predicates:
\[
E\vdash(\rho\land\sigma)(\tau)
\;\Longleftrightarrow\;
E\vdash\rho(\tau)\text{ and }E\vdash\sigma(\tau),
\]
\[
E\vdash(\rho\lor\sigma)(\tau)
\;\Longleftrightarrow\;
E\vdash\rho(\tau)\text{ or }E\vdash\sigma(\tau).
\]
We assume two properties of atoms. Evidence is \emph{sound} when
\(E\vdash R_k(\tau)\) implies \(R_k(\tau)\). It is \emph{monotone} when
\(E\subseteq E'\) and \(E\vdash R_k(\tau)\) imply \(E'\vdash R_k(\tau)\). Both
properties lift by induction to every positive predicate. By soundness, a
boundary that admits on \(E\vdash\rho_i(\tau_i)\) admits only transitions that
satisfy \(\rho_i\). By monotonicity, withholding evidence cannot produce an
admission that full evidence would deny. Negation breaks this for atoms of the
second kind: supporting \(\lnot R_k(\tau)\) requires evidence of absence, which
an open world does not provide, and reading ``not supported'' as ``false''
would let missing evidence cause admission.

The positive fragment is therefore the conservative choice, safe for both kinds
of atom. Negation is not excluded on mathematical grounds. With \(\lnot\)
added and formulas still identified extensionally, semantic implication and
the attenuation arguments can be retained, and the generated semantic algebra
is Boolean. The positive-DNF containment criterion of
Lemma~\ref{lem:dnf}, however, does not extend unchanged to formulas with
negation. A deployment may admit \(\lnot\) on atoms decidable at the boundary,
or under a closed-world assumption, but must use an appropriate implication
check. The validity conditions below are stated at the level of truth; a
boundary discharges them through sound evidence.

Conjunction and disjunction have their ordinary set semantics:
\[
\sem{\rho\land\sigma}
=
\sem{\rho}\cap\sem{\sigma},
\]
and
\[
\sem{\rho\lor\sigma}
=
\sem{\rho}\cup\sem{\sigma}.
\]

Thus one causal execution may require, for example,
\[
R_1\land R_2,
\]
another may allow
\[
R_1\lor R_2,
\]
and a later continuation may refine the condition to
\[
(R_1\lor R_2)\land R_3.
\]

\section{The Relational Predicate Order}

We now define the order that will carry relational continuity.

\begin{definition}[Relational refinement]
For relational predicates \(\rho'\) and \(\rho\), define
\[
\rho'\preceq_R\rho
\]
iff
\[
\rho'\Rightarrow\rho,
\]
that is, \(\rho'(\tau)=1\) implies \(\rho(\tau)=1\) for every
\(\tau\in\Transitions_J\). Equivalently,
\[
\sem{\rho'}
\subseteq
\sem{\rho}.
\]
\end{definition}

Throughout, \(\Rightarrow\) between predicates denotes this extensional
implication, under the fixed interpretation of the atoms.

A stronger predicate therefore denotes a smaller set of admissible transitions.

For example,
\[
R_1
\Rightarrow
R_1\lor R_2,
\]
so
\[
R_1
\preceq_R
R_1\lor R_2.
\]

Likewise,
\[
R_1\land R_3
\Rightarrow
R_1,
\]
so
\[
R_1\land R_3
\preceq_R
R_1.
\]

Hence the sequence
\[
R_1\lor R_2
\;\succeq_R\;
R_1
\;\succeq_R\;
R_1\land R_3
\]
is monotonically restrictive.

\begin{proposition}[Relational predicate lattice]\label{prop:lattice}
Identify positive relational predicates over \(\Transitions_J\) that have the
same semantic extension. Ordered by \(\preceq_R\), the resulting classes form a
distributive lattice with meet \(\land\) and join \(\lor\).
\end{proposition}

\begin{proof}
Under the semantic map \(\rho\mapsto\sem{\rho}\), conjunction maps to set
intersection and disjunction maps to set union. The extensions of positive
predicates are therefore closed under \(\cap\) and \(\cup\), and form a
sublattice of the powerset of \(\Transitions_J\) ordered by inclusion. Every
sublattice of a distributive lattice is distributive.
\end{proof}

A second identification is possible: the free positive algebra, in which
formulas are identified when they are equivalent under \emph{every}
interpretation of the atoms. It is finer than the extensional one, because two
atoms may have the same extension without being propositionally equivalent. The
closure results below use only the extensional order. The syntactic check that
follows is the interpretation-independent one.

This lattice is the formal version of a relational authorization space. A
successor may move downward by refinement, thereby accepting fewer relational
transitions, but may not move upward without an explicit re-authorization or a
new origin.

\paragraph{Checking refinement.}\label{sec:refinement}
As a semantic condition, \(\rho'\Rightarrow\rho\) depends on the extensions of
the atoms. A verifier need not compute them, because an
interpretation-independent refinement can be checked syntactically. Every
formula in the positive grammar is finite and can be written in disjunctive
normal form. Regard each conjunction as the finite set of atoms occurring in
it.

\begin{lemma}[Positive-DNF implication criterion]\label{lem:dnf}
Let
\[
\rho'=\bigvee_{a\in A}\bigwedge_{p\in P_a}p,
\qquad
\rho=\bigvee_{b\in B}\bigwedge_{p\in Q_b}p
\]
be positive formulas in DNF. Then \(\rho'\Rightarrow\rho\) holds under every
Boolean interpretation of the atoms iff
\[
\forall a\in A\;\exists b\in B:\quad Q_b\subseteq P_a.
\]
Equivalently, every disjunct of \(\rho'\) contains all atoms of some disjunct
of \(\rho\).
\end{lemma}

\begin{proof}
If \(Q_b\subseteq P_a\), every valuation satisfying the conjunction indexed by
\(P_a\) also satisfies the one indexed by \(Q_b\). The displayed condition
therefore makes every disjunct of \(\rho'\) imply \(\rho\), and hence makes
\(\rho'\Rightarrow\rho\) valid under every interpretation.

Conversely, suppose that for some \(a\in A\), no \(Q_b\) is contained in
\(P_a\). Assign true exactly to the atoms in \(P_a\) and false to every other
atom occurring in either formula. This valuation satisfies the \(a\)-th
disjunct of \(\rho'\). For every \(b\in B\), choose
\(p_b\in Q_b\setminus P_a\); that atom is false, so no disjunct of \(\rho\) is
satisfied. Thus \(\rho'\not\Rightarrow\rho\), proving the contrapositive.
\end{proof}

In particular, conjoining a further predicate
(\(\rho'=\rho\land\sigma\)) and deleting disjuncts are always refinements. The
criterion of Lemma~\ref{lem:dnf} is sound and decidable from the formulas alone.
It is conservative for the fixed semantic interpretation used by the model:
particular atom extensions may make \(\rho'\Rightarrow\rho\) hold even when the
interpretation-independent criterion fails.

\section{Relational Continuity Along a Causal Chain}

\subsection{Relational envelopes}

Associate with every execution step \(s_i\) both:

\begin{enumerate}
    \item an authority context
    \[
    C_i\subseteq O\times R,
    \]
    and
    \item a relational envelope
    \[
    \rho_i\in\Pred(\Transitions_J).
    \]
\end{enumerate}

The relational envelope specifies the class of executor-profile transitions
that remain admissible at that point in the lineage.

\begin{definition}[Authorization-aware causal history]\label{def:lineage}
Let \(\Lineages\) be the set of finite nonempty sequences
\[
\widehat{\ell}_i
:=
\langle
(s_0,C_0,\rho_0),\ldots,(s_i,C_i,\rho_i)
\rangle,
\]
where \(s_k\in S\), \(C_k\subseteq O\times R\), and
\(\rho_k\in\Pred(\Transitions_J)\). These histories are
\emph{authorization-aware}: their endpoint contains the state carried at that
occurrence. Write
\(\widehat{\ell}_i\preceq_L\widehat{\ell}_j\) when the former sequence is a
prefix of the latter. The index records a position inside this particular
history; it is not a global clock and does not identify an occurrence without
the history.
\end{definition}

\subsection{Extended valid transition}

\begin{definition}[Relationally valid transition]\label{def:valid-transition}
Let
\[
x_i=\exec(s_i),
\qquad
x_{i+1}=\exec(s_{i+1}),
\]
and
\[
\tau_i=
\big(
q_J(x_i),
q_J(x_{i+1})
\big).
\]
The transition from
\[
(s_i,C_i,\rho_i)
\]
to
\[
(s_{i+1},C_{i+1},\rho_{i+1})
\]
is relationally valid iff
\[
\begin{aligned}
\ValidR
(s_i,C_i,\rho_i,s_{i+1},C_{i+1},\rho_{i+1})
\Longleftrightarrow\;&
\PoR(s_i,s_{i+1})
\\
&\land\; C_{i+1}\subseteq C_i
\\
&\land\; \rho_i(\tau_i)
\\
&\land\;(\rho_{i+1}\Rightarrow\rho_i).
\end{aligned}
\]
\end{definition}

The four conjuncts have different roles:

\begin{enumerate}
    \item \(\PoR(s_i,s_{i+1})\) establishes causal continuation;
    \item \(C_{i+1}\subseteq C_i\) prevents authority expansion;
    \item \(\rho_i(\tau_i)\) requires the concrete executor transition to
    satisfy the currently admitted relation; and
    \item \(\rho_{i+1}\Rightarrow\rho_i\) prevents relational expansion.
\end{enumerate}

\begin{definition}[PIC-compatible realization]
\label{def:pic-compatible}
A concrete authorization system is \emph{PIC-compatible} with respect to an
authorization-aware lineage model when every admitted non-origin transition
\(s_i\to s_{i+1}\) is represented by endpoint states
\((s_i,C_i,\rho_i)\) and \((s_{i+1},C_{i+1},\rho_{i+1})\), indexed by the
corresponding causal-history prefixes, and the system:
\begin{enumerate}
    \item soundly establishes \(\PoR(s_i,s_{i+1})\);
    \item enforces \(C_{i+1}\subseteq C_i\);
    \item soundly establishes the profile facts used in
    \(\tau_i=(q_J(x_i),q_J(x_{i+1}))\) and admits only when
    \(\rho_i(\tau_i)\) is established, either by direct evaluation on
    verified profile facts or by sound evidence in the sense of
    Section~\ref{sec:relpred}; and
    \item enforces \(\rho_{i+1}\Rightarrow\rho_i\).
\end{enumerate}
Thus the admitted transitions realize Definition~\ref{def:valid-transition},
independently of which artifacts or policy engine carry and check its premises.
This definition classifies transition admission only. It does not imply that
every real-world effect is mediated or attributed to the correct lineage;
those are the separate premises of Assumptions~\ref{ass:mediation}
and~\ref{ass:attribution}.
\end{definition}

Two design points concern the third conjunct of
Definition~\ref{def:valid-transition}, \(\rho_i(\tau_i)\). First, hop \(i\) is governed by
the predecessor's envelope \(\rho_i\), not by the successor's \(\rho_{i+1}\).
The envelope carried at \(s_i\) constrains the transition that continues
\(s_i\); the envelope \(\rho_{i+1}\) constrains the transitions that will
continue \(s_{i+1}\). Requiring \(\rho_{i+1}(\tau_i)\) as well would impose a
stronger, retroactive policy that the model does not need.

Second, \(\tau_i\) refers to the class of \(x_{i+1}\), an executor that may not
exist, or may not yet be selected, when \(s_i\) emits its continuation. The
condition \(\rho_i(\tau_i)\) is therefore evaluated by the receiving side at
acceptance time, not by the predecessor at emission time. This is the temporal
gap that Proof-of-Continuity addresses for authority~\cite{gallo2026poc},
applied to the envelope: the predecessor fixes the envelope, and any successor
that later appears is admitted only if it falls inside it. The PIC project
documents the corresponding engineering motivation as the N+1 unknown-executor
problem~\cite{picNPlus1Executor}.

PoR supplies the causal spine. The authority and relational contexts are the two
monotone payloads propagated along that spine.

\section{Dual Monotonicity}

\begin{lemma}[Dual Monotonicity]
Let
\[
\pi=
\langle
(s_0,C_0,\rho_0),
\ldots,
(s_n,C_n,\rho_n)
\rangle
\]
be a chain in which every adjacent transition is relationally valid. Then
\[
C_n\subseteq C_{n-1}\subseteq\cdots\subseteq C_0
\]
and
\[
\rho_n
\preceq_R
\rho_{n-1}
\preceq_R
\cdots
\preceq_R
\rho_0.
\]
In particular,
\[
C_n\subseteq C_0
\]
and
\[
\rho_n\Rightarrow\rho_0.
\]
\end{lemma}

\begin{proof}
Every valid transition satisfies
\[
C_{i+1}\subseteq C_i.
\]
Transitivity of set inclusion gives
\[
C_n\subseteq C_0.
\]

Every valid transition also satisfies
\[
\rho_{i+1}\Rightarrow\rho_i.
\]
Transitivity of logical implication gives
\[
\rho_n\Rightarrow\rho_0.
\]
\end{proof}

The two monotonicities decrease in different mathematical spaces:

\[
C_i
\]
attenuates under set inclusion, while
\[
\rho_i
\]
attenuates under logical implication.

They nevertheless express the same continuity principle:

\begin{quote}
A valid successor may preserve or restrict what its predecessor allowed, but it
may not expand it.
\end{quote}

The proof is short by design. Validity imposes monotonicity at each hop, and
the lemma is its transitive closure, so the result is close to its definition.
The substance of the model lies elsewhere: in the choice of the two ordered
spaces, in their composition with PoR along one lineage, in the reading of the
earlier projection result as a quotient instance, and in the witness
construction of Section~\ref{sec:witness}, which conditionally realizes the
premises under its explicit assumptions.

\section{Relational Closure}

Dual monotonicity implies more than endpoint restriction. Every concrete
transition in the lineage remains inside the relational envelope established at
the origin.

\begin{theorem}[Origin Relational Closure]
Let
\[
\tau_i=
\big(
q_J(x_i),
q_J(x_{i+1})
\big)
\]
be the relational observation of hop \(i\). If every adjacent transition is
relationally valid, then
\[
\tau_i\in\sem{\rho_0}
\]
for every \(i=0,\ldots,n-1\).
\end{theorem}

\begin{proof}
Validity gives
\[
\rho_i(\tau_i)=1,
\]
hence
\[
\tau_i\in\sem{\rho_i}.
\]
By Dual Monotonicity,
\[
\rho_i\Rightarrow\rho_0,
\]
so
\[
\sem{\rho_i}\subseteq\sem{\rho_0}.
\]
Therefore
\[
\tau_i\in\sem{\rho_0}.
\]
\end{proof}

Thus a valid descendant cannot introduce an executor relation that the origin
relational envelope excluded.

\begin{corollary}[No relational expansion]
If
\[
\tau\notin\sem{\rho_0},
\]
then \(\tau\) cannot occur as a valid transition anywhere in the same
relationally continuous lineage.
\end{corollary}

\section{The Product Authorization State}

The two monotone dimensions can be represented as one ordered state.

\begin{definition}[Relational authorization state]
For the history of Definition~\ref{def:lineage}, define
\[
\Gamma(\widehat{\ell}_i)
:=
(C_i,[\rho_i]_{\mathrm{sem}}),
\]
and abbreviate it by \(\Gamma_i\). This is well-defined because the endpoint
tuple of \(\widehat{\ell}_i\) contains \(C_i\) and \(\rho_i\). The predicate
coordinate is its semantic-equivalence class from
Proposition~\ref{prop:lattice}; replacing \(\rho_i\) by a predicate with the
same extension does not change the state.
\end{definition}

\begin{definition}[Product attenuation order]
For states
\[
\Gamma'=(C',\rho')
\qquad\text{and}\qquad
\Gamma=(C,\rho),
\]
define
\[
\Gamma'\preceq\Gamma
\]
iff
\[
C'\subseteq C
\qquad\text{and}\qquad
\rho'\Rightarrow\rho.
\]
\end{definition}

\begin{proposition}[The product order is a partial order]
On
\[
\mathcal P(O\times R)\times
\big(\Pred(\Transitions_J)/{\equiv_{\mathrm{sem}}}\big),
\]
where \(\rho\equiv_{\mathrm{sem}}\sigma\) iff
\(\sem{\rho}=\sem{\sigma}\), the relation \(\preceq\) is a partial order. If
predicates are retained as syntactic formulas instead, the same definition is
only a preorder.
\end{proposition}

\begin{proof}
Reflexivity and transitivity follow componentwise from set inclusion and
logical implication. If both \((C',\rho')\preceq(C,\rho)\) and
\((C,\rho)\preceq(C',\rho')\), then \(C'=C\) and
\(\sem{\rho'}=\sem{\rho}\); hence the two states are equal in the displayed
quotient. Without quotienting formulas by semantic equivalence, distinct but
equivalent formulas witness the failure of antisymmetry.
\end{proof}

Every valid transition therefore satisfies
\[
\Gamma_{i+1}\preceq\Gamma_i.
\]

Hence:
\[
\Gamma_n
\preceq
\Gamma_{n-1}
\preceq
\cdots
\preceq
\Gamma_0.
\]

\begin{lemma}[Combined Closure]
For every valid descendant state
\(\Gamma_i=(C_i,[\rho_i]_{\mathrm{sem}})\),
\[
C_i\subseteq C_0
\]
and
\[
\sem{\rho_i}\subseteq\sem{\rho_0}.
\]
\end{lemma}

\begin{proof}
Immediate from Dual Monotonicity and the semantic equivalence
\[
\rho_i\Rightarrow\rho_0
\Longleftrightarrow
\sem{\rho_i}\subseteq\sem{\rho_0}.
\]
\end{proof}

The origin therefore bounds both:

\[
\boxed{\text{which privileges may continue}}
\]
and
\[
\boxed{
\begin{gathered}
\text{through which executor relationships}\\
\text{they may continue}
\end{gathered}
}.
\]

\section{Causal History Grows While Authorization Freedom Shrinks}

The lineage itself is not one of the shrinking objects. Under the prefix order
of Definition~\ref{def:lineage},
\[
\widehat{\ell}_0
\preceq_L
\widehat{\ell}_1
\preceq_L
\cdots
\preceq_L
\widehat{\ell}_n.
\]
The temporal axis here is causal prefix order, not wall-clock time: timestamps,
expiry and elapsed time may be evidence or policy inputs, but none determines
the position \(\widehat{\ell}_i\) in the causal execution history.

Thus the three dimensions move together but not identically:
\[
\begin{array}{ccccccc}
s_0 &\rightarrow& s_1&\rightarrow&s_2&\rightarrow&\cdots s_n\\[3pt]
C_0&\supseteq&C_1&\supseteq&C_2&\supseteq&\cdots C_n\\[3pt]
\rho_0&\Leftarrow&\rho_1&\Leftarrow&\rho_2&\Leftarrow&\cdots\rho_n\\[3pt]
\widehat{\ell}_0&\preceq_L&\widehat{\ell}_1&\preceq_L&
\widehat{\ell}_2&\preceq_L&\cdots\widehat{\ell}_n
\end{array}
\]
where
\[
\rho_{i+1}\Rightarrow\rho_i.
\]

More generally, if \(\widehat{\ell}_i\preceq_L\widehat{\ell}_j\) and every
intervening transition is relationally valid, Dual Monotonicity gives
\[
\Gamma(\widehat{\ell}_j)\preceq
\Gamma(\widehat{\ell}_i).
\]
The precise statement is therefore not that a lineage in \(\Lineages\) itself
decreases. Rather, causal history grows by prefix extension while the
authorization contexts \emph{carried along that history} attenuate in two
dimensions.

This yields the structural principle:
\[
\boxed{
\begin{gathered}
\text{causal history accumulates while}\\
\text{authorization freedom attenuates.}
\end{gathered}
}
\]

\section{The Confused Deputy Under Dual Continuity}

\subsection{Authority mixing}

Suppose an executor \(D\) processes a request whose origin context is \(C_0\)
while simultaneously holding an independent authority source \(C_D\). Let
\[
(o,r)\notin C_0
\qquad\text{but}\qquad
(o,r)\in C_D.
\]

A lineage-invariant possession model may authorize the action by selecting
\(C_D\), even though the request itself did not carry that authority. This is
the authority/causality mismatch that produces the confused deputy in the
multi-step setting~\cite{gallo2026poc}.

\subsection{Valid exercise}

The authorization rule of~\cite{gallo2026poc} links the exercise of a privilege
to the context carried at the step that exercises it. We restate it for the
relational model.

\begin{definition}[Valid exercise]\label{def:exercise}
Let
\[
\pi=
\langle
(s_0,C_0,\rho_0),
\ldots,
(s_n,C_n,\rho_n)
\rangle
\]
be a chain. A step \(s_i\) of \(\pi\) \emph{validly exercises} \((o,r)\) iff
\[
(o,r)\in C_i
\]
and either \(i=0\), or the transition from \(s_{i-1}\) to \(s_i\) is
relationally valid; in the latter case \(\rho_{i-1}(\tau_{i-1})=1\) in
particular.
\end{definition}

The relational condition is indexed by the hop that enters \(s_i\), because an
envelope governs the transition that continues the step carrying it.

\begin{proposition}[No in-lineage exercise beyond the origin]
\label{prop:inlineage}
If every transition of \(\pi\) up to \(s_i\) is relationally valid and \(s_i\)
validly exercises \((o,r)\), then
\[
(o,r)\in C_0.
\]
\end{proposition}

\begin{proof}
By Definition~\ref{def:exercise}, \((o,r)\in C_i\). The prefix
\(\langle(s_0,C_0,\rho_0),\ldots,(s_i,C_i,\rho_i)\rangle\) is itself a chain
whose adjacent transitions are all relationally valid, so Combined Closure
applies to it and gives \(C_i\subseteq C_0\). Hence \((o,r)\in C_0\).
\end{proof}

\paragraph{Interpretation: no in-lineage authority mixing.}
Even if \(\exec(s_i)=D\) and \(D\) independently holds authority for \((o,r)\)
from a source outside the lineage, that fact does not inject \((o,r)\) into
\(C_i\). The model neither identifies \(C_i\) with all authority held by
\(\exec(s_i)\) nor supplies an update rule that adds independently held
privileges to \(C_{i-1}\). Such an update would be an authority expansion unless
the added privileges were already in \(C_{i-1}\), whereas validity requires
\(C_i\subseteq C_{i-1}\). Thus independent authority may exist while remaining
semantically external to the current lineage:
\[
\boxed{
\begin{gathered}
(o,r)\notin C_0\\
\Longrightarrow\\
(o,r)\text{ cannot be a valid}\\
\text{in-lineage exercise}.
\end{gathered}
}
\]
This conclusion is independent of executor identity. Different steps may be
realized by different executors, but authority follows the execution lineage,
not the union of their ambient privilege sets.

For example, let \(C_0(req)=\{(\operatorname{read},A)\}\), while deputy \(D\)
independently holds \((\operatorname{write},B)\). Even if
\(\exec(s_i)=D\), validity gives \(C_i\subseteq C_0(req)\), hence
\((\operatorname{write},B)\notin C_i\). The deputy possesses the authority; the
execution does not. This is the precise sense in which
\[
\boxed{
\begin{gathered}
\text{PIC validity}\\
\Longrightarrow\\
\text{no in-lineage confused-deputy}\\
\text{authority mixing}.
\end{gathered}
}
\]
The result is internal to the PIC semantics. It depends only on the
lineage-local authorization state and relational validity; it does not identify
\(C_i\) with all ambient authority held by \(\exec(s_i)\), and it does not
require Complete Mediation or Causal Attribution.

Proposition~\ref{prop:inlineage}, however, only constrains steps \emph{of} the
lineage. It does not by itself establish that every real-world effect caused by
a request is presented in that lineage. In particular, a deputy may perform an
action outside it by opening a lineage of its own, rooted in its own privileges
on the target the request designated. This is Hardy's compiler, which writes
the billing file with its own authority as if it were
housekeeping~\cite{hardy1988confused}. The previous paper identifies this gap
as lineage origination and leaves it to future work~\cite{gallo2026poc}.
Closing it requires the additional premises stated below. In compact form, the
global result has the stronger premise:
\[
\boxed{
\begin{gathered}
\text{PIC validity + Complete Mediation}\\
\text{+ Causal Attribution}\\
\Downarrow\\
\text{no request-caused accepted effect}\\
\text{outside the origin context}.
\end{gathered}
}
\]

\subsection{Exclusion of the confused deputy}

We first make the objects used by the result explicit. Let \(\mathsf{Req}\) be
the set of requests and \(\mathsf{Eff}\) the set of modeled security-relevant
effects under consideration. Every
request has an authorization-aware lineage through a map
\[
\operatorname{ReqLineage}:\mathsf{Req}\rightarrow\Lineages,
\qquad
req\longmapsto\widehat{\ell}_{req},
\]
and \(C_0(req)\) is the authority context in the first tuple of
\(\widehat{\ell}_{req}\). Let
\[
\operatorname{step}:\mathsf{Eff}\rightarrow S,
\qquad
\operatorname{realize}:\mathsf{Eff}\rightarrow O\times R,
\]
associate an effect with the step that performs it and the privilege it
realizes. Let
\[
\operatorname{Lineage}:\mathsf{Eff}\rightarrow\Lineages
\]
be the lineage to which the enforcement system attributes the effect, and let
\[
\operatorname{CausedBy}\subseteq\mathsf{Eff}\times\mathsf{Req}
\]
be the semantic causal relation between effects and requests. Finally, write
\(\operatorname{IndependentAuthority}(D,(o,r))\) when executor \(D\) holds
authority for \((o,r)\) from a source other than the request context. An effect
\(e\) is performed by \(D\) when
\(\exec(\operatorname{step}(e))=D\).

The distinction between \(\operatorname{CausedBy}\) and
\(\operatorname{Lineage}\) is intentional. The former is the causal fact that
the security statement concerns; the latter is the attribution made available
to enforcement.

\begin{definition}[Confused deputy, lineage form]\label{def:cd}
A confused deputy occurs iff there exist
\(req\in\mathsf{Req}\), \(e\in\mathsf{Eff}\), \(D\in X\), and
\((o,r)\in O\times R\) such that
\[
\begin{aligned}
&\operatorname{CausedBy}(e,req),
&&\operatorname{realize}(e)=(o,r),\\
&\exec(\operatorname{step}(e))=D,
&&(o,r)\notin C_0(req),\\
&\operatorname{IndependentAuthority}(D,(o,r)).
\end{aligned}
\]
\end{definition}

The previous paper states its privilege-exclusion condition as
\((o,r)\notin Priv(U)\) for the originating principal \(U\). If the origin gate
selects \(C_0(req)\subseteq Priv(U)\), then that condition implies
\((o,r)\notin C_0(req)\). Under this explicit origin-selection premise, every
confused deputy in the sense of~\cite{gallo2026poc} is therefore one in the
sense of Definition~\ref{def:cd}. This comparison is not used by the theorem
below, whose abstract initial context is simply \(C_0(req)\). The lineage form
is the stronger target when the premise holds: it also covers privileges that
the originating principal holds but deliberately left out of \(C_0(req)\).

\begin{definition}[Accepted effect]\label{def:accepted}
For \(\widehat{\ell}\in\Lineages\), an effect \(e\) is \emph{accepted} in
\(\widehat{\ell}\) for \((o,r)\), written
\(\operatorname{Accepted}(e,\widehat{\ell},(o,r))\), iff
\(\operatorname{realize}(e)=(o,r)\),
\(\operatorname{step}(e)=s_k\) for some tuple \((s_k,C_k,\rho_k)\) of
\(\widehat{\ell}\), every transition in the prefix ending at that tuple is
relationally valid, and \(s_k\) validly exercises \((o,r)\).
\end{definition}

Here \(\operatorname{realize}(e)\) denotes the operation and actual resource on
which the effect occurs after name resolution, aliasing, redirection and other
designation processing. Checking only an input string, pathname or logical
designation is insufficient if it may resolve to a different resource. Complete
mediation requires the verifier to authorize the realized pair \((o,r)\).
Predecessor binding and destination binding are therefore separate: the
checkpoint binds a continuation to its predecessor and carried state, while the
complete mediator binds authorization to the operation and resource actually
realized. The checkpoint alone does not make an unresolved designation an
authorized destination.

\begin{assumption}[Complete mediation]\label{ass:mediation}
Every occurring effect is accepted under its attributed lineage:
\[
\forall e\in\mathsf{Eff},\qquad
\operatorname{Accepted}\bigl(
e,\operatorname{Lineage}(e),\operatorname{realize}(e)
\bigr).
\]
No path to the affected resource bypasses the verifier.
\end{assumption}

\begin{assumption}[Causal attribution]\label{ass:attribution}
Every effect causally triggered by a request is presented as a step of that
request's lineage:
\[
\begin{aligned}
&\forall e\in\mathsf{Eff}\;\forall req\in\mathsf{Req},\\
&\operatorname{CausedBy}(e,req)
\Longrightarrow
\operatorname{Lineage}(e)=\widehat{\ell}_{req}.
\end{aligned}
\]
\end{assumption}

A third property is related but distinct.

\begin{assumption}[Origination control]\label{ass:origination}
A new lineage is opened only at an origin gate independent of every executor,
and executors hold no authority of their own outside lineages opened there.
\end{assumption}

Origination control and causal attribution are not equivalent. Origination
control restricts \emph{who} may open a lineage, but the gate may still be
invoked because of a request, and would then re-root a request-caused effect in
a new lineage: origination control holds while attribution fails. Conversely,
attribution can hold while executors open genuinely independent lineages of
their own. The theorem below uses complete mediation and causal attribution;
origination control is one ingredient of how an architecture realizes them.

\begin{theorem}[No request-caused effect beyond the origin]\label{thm:causal-bound}
Under Assumptions~\ref{ass:mediation} and~\ref{ass:attribution}, with the
authority-attenuation clause \(C_{i+1}\subseteq C_i\) built into relational
validity, for every
\(req\in\mathsf{Req}\), \(e\in\mathsf{Eff}\), and \((o,r)\in O\times R\),
\[
\begin{aligned}
&\operatorname{CausedBy}(e,req)
\land
\operatorname{realize}(e)=(o,r)\\
&\hspace{2em}\Longrightarrow
(o,r)\in C_0(req).
\end{aligned}
\]
\end{theorem}

\begin{proof}
Suppose an effect \(e\) caused by \(req\) realizes \((o,r)\). By
Assumption~\ref{ass:attribution}, \(e\) is presented as a step \(s_k\) of
\(\widehat{\ell}_{req}\). By Assumption~\ref{ass:mediation},
\(\operatorname{Accepted}(e,\widehat{\ell}_{req},(o,r))\) holds, so every
transition of \(\widehat{\ell}_{req}\) up to \(s_k\) is relationally valid and \(s_k\) validly
exercises \((o,r)\). Applying Proposition~\ref{prop:inlineage} to the
prefix of \(\widehat{\ell}_{req}\) ending at \(s_k\) gives \((o,r)\in C_0\).
Here \(C_0=C_0(req)\), the origin context of
\(\widehat{\ell}_{req}\).
\end{proof}

\begin{corollary}[Exclusion of the confused deputy]\label{thm:cd}
Under Assumptions~\ref{ass:mediation} and~\ref{ass:attribution}, the
confused-deputy condition of Definition~\ref{def:cd} is unsatisfiable. Under
the additional origin-selection premise
\(C_0(req)\subseteq Priv(U)\) stated above, the condition
of~\cite{gallo2026poc} is unsatisfiable as well.
\end{corollary}

\begin{proof}
Theorem~\ref{thm:causal-bound} contradicts
\((o,r)\notin C_0(req)\). The independent-authority conjunct is not needed;
the theorem proves the stronger statement that \emph{no} request-caused effect,
whether or not its executor independently holds the privilege, can exercise
authority outside the request's origin context.
Under the additional origin-selection premise, the comparison preceding
Definition~\ref{def:accepted} shows that every confused deputy in the sense
of~\cite{gallo2026poc} satisfies Definition~\ref{def:cd}, so that condition is
unsatisfiable as well.
\end{proof}

The contradiction is direct, and the enforcement properties that the proof
does not derive sit in two named assumptions. The monotone-authority premise
enters through relational validity and Proposition~\ref{prop:inlineage}; the
relational-envelope premise is not needed for the authority conclusion.
The result does not say that a deputy cannot contain a logic error or attempt an
operation outside \(C_0(req)\). It says that, under complete mediation and
causal attribution, such an error cannot produce an \emph{accepted}
request-caused effect exercising authority outside that context.
The deputy may possess the independent authority; possession does not make that
authority admissible in the request's causal history.

\paragraph{What the construction realizes, and what it does not.}
The boundary-witnessed construction of Section~\ref{sec:witness} can discharge
complete mediation for resources whose every access is forced through its
boundaries; exclusive resource mediation is a deployment property, not a
consequence of the checkpoint syntax. Under its Witness Assumptions, the
construction discharges causal attribution for \emph{explicit mediated
handoffs}: a checkpoint is issued only for a step observed consuming the
continuation of a verified predecessor.
For resources subject to that exclusive mediation, origination control also
prevents an executor from opening an independently privileged lineage outside
the gate.

It does not realize attribution for causality that flows through state. A
request may modify an executor's memory, a queue, a database, a cache, a file
or shared memory; the request may then complete, and a later timer, asynchronous
callback or background worker may read that state and obtain a new lineage at
the origin gate. The resulting effect depends causally on the request without
consuming its continuation, and no handoff observation will attribute it.
Designation that flows as data is the typical case: a deputy performs
housekeeping on a target it read from the request. Indirect signaling, covert
channels and side channels can likewise influence control or data flow without
appearing as an explicit mediated handoff. Information-flow tracking or a
stronger execution model can address modeled overt channels, but ordinary
tracking does not by itself establish attribution for every covert or side
channel. Otherwise a further causal-attribution assumption is required, for
example that executors retain no state across requests. The attribution itself
therefore belongs to the trusted computing base of the enforcement
architecture. Corollary~\ref{thm:cd} follows deductively from its two premises;
the construction discharges attribution for mediated handoffs, and not beyond.

The canonical Hardy case falls precisely within this limitation when a
request-supplied designation influences an effect through the deputy's
independent authority without appearing as an explicit mediated handoff. The
effect is semantically request-caused, but the witness construction alone may
not attribute it to the request history. Exclusion of that case therefore
remains conditional on Causal Attribution and Complete Mediation; this is a
limit of the construction, not of the conditional theorem.
PIC eliminates the authority-mixing step once an effect is inside the request's
valid lineage. The remaining deployment problem is ensuring that
request-caused effects cannot escape attribution to that lineage.

\paragraph{Exclusive mediated ingress.}
A strongly confined deployment may impose the sufficient operational condition
that every modeled security-relevant external input capable of influencing a
request-dependent effect enters as a boundary-mediated continuation associated
with a lineage. Combined with complete mediation of effects and
lineage-preserving treatment of every modeled internal channel, this gives the
verifier an ingress and egress point for all such influence. This is a
deployment condition, not a premise of the core closure results, and it does not
establish attribution for unmodeled covert or side channels.

The relational component is not needed for this result. Its contribution is
orthogonal: it closes the class of admissible executor relationships at the
same time.

\begin{corollary}[Dual closure against authority and relationship import]
In a relationally continuous lineage:
\begin{enumerate}
    \item no step validly exercises, and no state carries, a privilege absent
    from \(C_0\); and
    \item no hop has a transition profile outside \(\sem{\rho_0}\), and no
    envelope carried along the lineage admits one.
\end{enumerate}
\end{corollary}

This is the strengthened closure result. Authority cannot be imported from an
unrelated source, and relational admissibility cannot be widened in order to
make an otherwise inadmissible continuation appear valid.

\section{Authorization Without Successor-Side Proof of Possession}

We now separate the semantic authorization rule from the mechanisms used to
produce evidence for that rule.

\subsection{Authentication and authorization are separate layers}

Fixing the steps also fixes their actual executors, since
\(x_i=\exec(s_i)\). To isolate the executor-dependent input, define the
following counterfactual decision template for fixed
\(s_i,C_i,\rho_i,s_{i+1},C_{i+1},\rho_{i+1}\):
\(V:X\times X\rightarrow\{0,1\}\) by
\[
\begin{aligned}
V(x,x')
:={}&
\PoR(s_i,s_{i+1})
\land C_{i+1}\subseteq C_i\\
&{}\land\rho_i\big(q_J(x),q_J(x')\big)\\
&{}\land(\rho_{i+1}\Rightarrow\rho_i),
\end{aligned}
\]
so that
\[
\begin{aligned}
&\ValidR(s_i,C_i,\rho_i,\\
&\qquad s_{i+1},C_{i+1},\rho_{i+1})\\
&\hspace{3em}=V(x_i,x_{i+1}).
\end{aligned}
\]
\(\PoR\) is a condition on the fixed steps, not on the two arguments of this
template. Thus \(V\) is used only to expose factorization of the relational
input; it does not assert that changing an executor while holding a step fixed
produces another execution admitted by the model.

\begin{proposition}[Relational executor dependence factors through the profile]
\label{prop:layers}
Conditional on fixed execution occurrences and authorization state, the
executor-dependent relational component of the decision factors through
\(q_J\times q_J\): the function \(V\) is invariant under \(\equiv_J\) in each
argument, and therefore factors uniquely as
\[
V=\bar V\circ(q_J\times q_J),
\qquad
\bar V:\Transitions_J\rightarrow\{0,1\}.
\]
\end{proposition}

\begin{proof}
The executors occur in \(V\) only through
\((q_J(x),q_J(x'))\). If \(x\equiv_J y\) and \(x'\equiv_J y'\), then
\(q_J(x)=q_J(y)\) and \(q_J(x')=q_J(y')\), so \(V(x,x')=V(y,y')\). Apply
Proposition~\ref{prop:factorization} to \(Y=X\times X\) with the product
equivalence \(\equiv_J\times\equiv_J\). Its quotient is in canonical bijection
with \(\Profiles_J\times\Profiles_J=\Transitions_J\) via \(q_J\times q_J\), which
is surjective onto \(\Transitions_J\).
\end{proof}

The proposition is deliberately modest. It does not say that the executor is
irrelevant to the decision: the occurrences determine their executors, and
\(\PoR\) depends on the occurrences. By itself it proves invariance under
executor replacement once the profiles and the fixed occurrence-level data are
given; it does not quantify over authentication mechanisms. We now state the
corresponding, and still deliberately limited, mechanism claim explicitly.

For the fixed occurrence and authorization data used above, define
\[
\begin{aligned}
D(u,v):={}&
\PoR(s_i,s_{i+1})
\land C_{i+1}\subseteq C_i\\
&{}\land\rho_i(u,v)
\land(\rho_{i+1}\Rightarrow\rho_i),\\
&(u,v)\in\Transitions_J.
\end{aligned}
\]

\begin{definition}[Sound profile-evidence mechanism]\label{def:profile-evidence}
A profile-evidence mechanism \(M\) consists of an evidence space
\(\mathcal E_M\) and a partial extraction function
\[
\alpha_M:X\times\mathcal E_M\rightharpoonup\Profiles_J.
\]
It is sound when
\[
\alpha_M(x,e)=u\Longrightarrow u=q_J(x)
\]
for every input on which it is defined. This definition concerns only evidence
for the executor profile; evidence for \(\PoR\) and for external relational
atoms remains subject to its own soundness assumptions.
\end{definition}

\begin{proposition}[Profile-preserving evidence-mechanism invariance]
\label{prop:mechanism-invariance}
Let \(M\) and \(N\) be profile-evidence mechanisms. If
\[
\begin{aligned}
\alpha_M(x,e)&=\alpha_N(y,f)=u,\\
\alpha_M(x',e')&=\alpha_N(y',f')=v,
\end{aligned}
\]
then both mechanisms induce the same authorization verdict \(D(u,v)\). In
particular, two sound mechanisms defined for the same concrete executors
\(x_i,x_{i+1}\) induce the same verdict.
\end{proposition}

\begin{proof}
The definition of \(D\) has no argument for a mechanism or for its evidence;
its executor-dependent input is exactly \((u,v)\). Equality of those inputs
therefore gives equality of verdicts. For sound mechanisms applied to the same
executors, Definition~\ref{def:profile-evidence} gives
\(u=q_J(x_i)\) and \(v=q_J(x_{i+1})\) for each mechanism.
\end{proof}

This proposition is an output-preservation result. It does not claim that all
mechanisms can produce evidence, that they have equal security properties, or
that the evidence mechanism for \(\PoR\) may be replaced without re-establishing
its soundness. It states precisely the separation used here: once a sound
mechanism establishes the same profile facts, its internal use or non-use of
PoP is not inspected by \(D\). Possession, where it is used, therefore
contributes evidence to the layer that establishes membership in a relational
class. Holding the same causal and authorization state fixed, the accepted
transition is the same; consequently, the closure results downstream of that
transition are unchanged. This does not transfer the security properties of one
evidence mechanism to another: each must separately satisfy its soundness
obligation. Authentication establishes the executor profiles, PoR establishes
causal continuation, \(C_i\) bounds the privileges that may continue, and
\(\rho_i\) bounds the profile transitions through which they may continue.

Membership is established at admission. The profile is fixed when the boundary
accepts the step and records it in the checkpoint of
Section~\ref{sec:witness}. ``Once authenticated'' therefore means for this
admission, not forever; how class membership changes over time, and how
revocation affects it, remain among the limitations of
Section~\ref{sec:limitations}.

Section~\ref{sec:picx} describes a realization that takes this layered route:
the successor authenticates with possession, and the decision is taken over
what that authentication established.

\subsection{Possession and relationship are different predicates}

A possession predicate has the general form
\[
\operatorname{Possess}(x,k),
\]
meaning that executor \(x\) controls artifact \(k\).

A Proof-of-Possession establishes some proposition of that form, or a closely
related holder-binding proposition:
\[
\PoP(x,k).
\]
It does not by itself establish \(q_J(x)=u\). Inferring profile membership
requires a trusted binding from \(k\) or its credential to the executor and to
the asserted class attributes. Conversely, a sound profile-evidence mechanism
may identify and classify the executor by boundary-local observation without
requiring successor-side PoP.

Nor does possession determine causal history. Two occurrences may have the
same executor, key, credential, authenticated profile and possession evidence
while belonging to different authorization-aware histories. Neither
\(\operatorname{Lineage}(e)=\widehat{\ell}\) nor
\(\PoR(s_i,s_{i+1})\) follows from \(\operatorname{Possess}(x,k)\).

A Proof of Relationship establishes something categorically different:
\[
\PoR(s_i,s_{i+1}),
\]
namely that one execution occurrence validly continues another.

The two propositions may be connected operationally, but neither is logically
identical to the other.

\subsection{Evidence is not semantics}

A concrete implementation may use a signature, key, credential, attestation,
trusted execution environment, authenticated channel, capability handoff, or
another mechanism to provide evidence for one of the following facts:

\[
q_J(x)=u,
\]
\[
\rho(\tau)=1,
\]
or
\[
\PoR(s_i,s_{i+1}).
\]

PoP may be one such evidence mechanism. For example, possession of a
predecessor-issued continuation artifact may witness a single causal handoff.
But the authorization semantics consumes the proved relationship and the
monotone state; it need not consume possession itself.

\begin{definition}[Possession-independent relational authorization]
\label{def:possindep}
A relational authorization rule is possession-independent when its logical
premises contain no predicate of the form
\[
\operatorname{Possess}(x,k)
\]
or
\[
\PoP(x,k).
\]
\end{definition}

\begin{proposition}[Semantic possession-independence]
\label{prop:semantic-possession-independence}
Once \(\PoR(s_i,s_{i+1})\), the executor profiles, the relational predicate
verdict and the two monotonicity conditions are given, the relational
authorization rule of Definition~\ref{def:valid-transition} is
possession-independent in the sense of Definition~\ref{def:possindep}.
\end{proposition}

\begin{proof}
The rule's premises are exactly
\(\PoR(s_i,s_{i+1})\), \(C_{i+1}\subseteq C_i\),
\(\rho_i(\tau_i)\), and \(\rho_{i+1}\Rightarrow\rho_i\). None is a predicate
of the form \(\operatorname{Possess}(x,k)\) or \(\PoP(x,k)\).
\end{proof}

This proposition records an immediate semantic fact about the abstract rule;
it does not show that its premises can be soundly realized without
successor-side PoP. That separate, substantive question is addressed by the
conditional construction below.

Proposition~\ref{prop:layers} places successor-side possession, where it is
used, in the authentication layer. The construction below shows that even that
layer can be discharged without it. This matters for the non-necessity claim:
the rule would be empty if every sound construction of PoR required the
successor to prove possession of something: a non-necessity result has content
only if its premises are satisfiable without the excluded predicate. We
therefore give a conditional construction in which the successor proves
possession of nothing, and state rather than conceal the environmental
assumptions on which it depends.

\subsection{A construction without successor-side possession}
\label{sec:witness}

Let \(h\) be a collision-resistant commitment to the state of an execution
step, and \(H\) a collision-resistant hash. Let each step \(s_i\) be accepted at
an enforcement boundary \(B_i\): the verifier that evaluates \(\ValidR\) for the
hop entering \(s_i\), and that holds a signing key. Write
\(\sigma_B(\operatorname{enc}(m))\) for a signature by \(B\) on the canonical
encoding of a structured payload \(m\), and let \(\lambda\) be a lineage
identifier fixed at the origin. All signed and hashed tuples use the same
injective canonical encoding \(\operatorname{enc}\).

A checkpoint is the tuple
\[
\kappa_i=
\bigl(B_i,m_i,\sigma_{B_i}(\operatorname{enc}(m_i))\bigr).
\]
It is identified by its signer and signed payload, not by the bytes of its
signature. Its identifier is
\[
\operatorname{cid}(\kappa_i)
:=
H\big(\operatorname{enc}(B_i,m_i)\big).
\]
Signature schemes may be malleable, admitting several valid signatures on one
payload; all of them have the same identifier, so malleability affects neither
predecessor binding nor the consumption register.

\begin{definition}[Boundary-witnessed checkpoint]
The checkpoint of step \(s_{i+1}\) is
\[
\kappa_{i+1}
:=
\bigl(
B_{i+1},m_{i+1},
\sigma_{B_{i+1}}(\operatorname{enc}(m_{i+1}))
\bigr),
\]
\[
m_{i+1}
:=
\left(
\begin{gathered}
\operatorname{cid}(\kappa_i),\,h(s_{i+1}),\,\lambda,\\
C_{i+1},\,\rho_{i+1},\,q_J(x_{i+1}),\,\nu_{i+1}
\end{gathered}
\right),
\]
where \(\nu_{i+1}\) is a nonce generated by \(B_{i+1}\). The boundary
\(B_{i+1}\) issues \(\kappa_{i+1}\) only after:
\begin{enumerate}
    \item verifying the predecessor checkpoint \(\kappa_i\) and that it carries
    the same lineage identifier \(\lambda\);
    \item checking that \(\operatorname{cid}(\kappa_i)\) has not already been
    consumed as the predecessor of an accepted step;
    \item observing that \(s_{i+1}\) consumes the continuation emitted by the
    step committed in \(\kappa_i\), written
    \(\operatorname{ObservedHandoff}(s_i,s_{i+1})\);
    \item identifying the executor \(x_{i+1}\) and determining its profile
    \(q_J(x_{i+1})\); and
    \item checking \(C_{i+1}\subseteq C_i\), obtaining
    \(E_{i+1}\vdash\rho_i(\tau_i)\), and checking
    \(\rho_{i+1}\Rightarrow\rho_i\), where \(E_{i+1}\) is the evidence
    available to the receiving boundary, and where \(C_i\), \(\rho_i\) and
    \(q_J(x_i)\) are read from \(\kappa_i\) and
    \(\tau_i=(q_J(x_i),q_J(x_{i+1}))\). Refinement is checked by the syntactic
    criterion of Section~\ref{sec:refinement}, which is sound.
\end{enumerate}
The origin checkpoint uses the same payload with a distinguished origin marker
\(\bot\) in place of \(\operatorname{cid}(\kappa_{-1})\):
\[
\begin{aligned}
m_0={}&(\bot,h(s_0),\lambda,C_0,\\
&\rho_0,q_J(x_0),\nu_0),\\
\kappa_0={}&\bigl(B_0,m_0,\\
&\sigma_{B_0}(\operatorname{enc}(m_0))\bigr).
\end{aligned}
\]
Authority bootstrap and the decision to issue \(\kappa_0\) are out of scope as
in~\cite{gallo2026poc}.
\end{definition}

Write \(\operatorname{stepCommitment}(\kappa_i)\) for the second component of
the checkpoint payload, namely \(h(s_i)\).

The nonce plays no challenge-response role: no successor answers it. It
distinguishes checkpoints issued by the same boundary for otherwise identical
steps. Since the identifier also covers the issuing boundary \(B_i\), entries in
the consumption register of the single-consumption/anti-replay assumption below
are unambiguous across boundaries as well.

The construction yields a predicate
\[
\PoR_\kappa(s_i,s_{i+1}),
\]
which holds iff \(\kappa_{i+1}\) verifies, its payload's first component is
\(\operatorname{cid}(\kappa_i)\) for a verified \(\kappa_i\), and both carry
\(\lambda\), with
\[
\begin{aligned}
\operatorname{stepCommitment}(\kappa_i)
&=h(s_i),\\
\operatorname{stepCommitment}(\kappa_{i+1})
&=h(s_{i+1}).
\end{aligned}
\]
This constructed predicate is distinct from the semantic \(\PoR(s_i,s_{i+1})\), which
holds iff \(s_{i+1}\) is a valid causal continuation of \(s_i\). What must be
shown is that \(\PoR_\kappa\) implies \(\PoR\).

Binding \(\operatorname{cid}(\kappa_i)\) rather than only the commitment
\(h(s_i)\) ties each checkpoint to the exact predecessor payload, and with it to the context
\(C_i\), the envelope \(\rho_i\), the profile \(q_J(x_i)\) and the lineage
\(\lambda\) that the checks in item~5 read. Except with negligible probability,
a checkpoint cannot be spliced onto a different predecessor that happens to
share a step commitment.

We work in the computational model. Collision resistance does not make \(H\)
injective: collisions exist, but no efficient adversary finds one except with
probability negligible in the security parameter \(\eta\). Statements about
uniqueness below therefore hold except with negligible probability, not
absolutely.

The construction rests on seven separate assumptions.

\begin{assumption}[Witness assumptions]\label{ass:witness}
\leavevmode
\begin{enumerate}
    \item \emph{Unforgeability and binding.} Except with probability negligible
    in \(\eta\), no efficient adversary forges a boundary signature or finds a
    collision in \(h\) or \(H\). Verifiers possess authentic verification keys
    for the boundaries they trust.
    \item \emph{Protocol compliance.} Boundaries follow the protocol: a
    boundary issues a checkpoint only after performing checks 1--5 correctly,
    and records each consumed predecessor.
    \item \emph{Handoff observation and adequacy.} A boundary observes, by its
    own means, which continuation a step consumes. Observed consumption is the
    operational criterion of causal continuation:
    \[
    \operatorname{ObservedHandoff}(s_i,s_{i+1})
    \Longrightarrow
    \PoR(s_i,s_{i+1}).
    \]
    \item \emph{Non-PoP profile observation.} A boundary identifies the
    executor it admits, and soundly determines its profile \(q_J\), from
    boundary-local observation or independently authenticated infrastructure
    evidence, without requiring the successor to demonstrate control of any
    artifact. For example, the boundary may hold a trusted mapping from the
    execution slot, container or virtual machine it controls to the executor,
    and from the executor to its profile:
    \[
    \text{slot}\;\longmapsto\;x_{i+1}\;\longmapsto\;q_J(x_{i+1}).
    \]
    \item \emph{Relational evidence soundness.} For every atom occurring in an
    admitted envelope, the boundary either computes the atom from verified
    profiles or obtains sound, monotone evidence for its truth in the sense of
    Section~\ref{sec:relpred}.
    \item \emph{Single-consumption / anti-replay.} A checkpoint identifier is
    accepted as the predecessor of at most one step, and no boundary reuses a
    nonce. This requires state: a consumption register on which acceptance and
    marking are atomic, shared by the boundaries that may receive a given
    continuation, or a continuation that designates a single verifier with an
    atomic local register.
    \item \emph{Root validity.} The origin gate issues \(\kappa_0\) only with
    \[
    \operatorname{stepCommitment}(\kappa_0)=h(s_0),
    \]
    a fresh lineage
    identifier and nonce, the origin contexts \(C_0,\rho_0\) selected by the
    bootstrap policy, and a sound profile \(q_J(x_0)\) for the executor of
    \(s_0\). The root tuple is canonically encoded and signed by an authenticated
    origin boundary. This assumes the truth of the root payload; a valid
    signature alone does not establish it.
\end{enumerate}
``By its own means'' means without relying on credentials presented by the
executor; the paragraph after the proof explains why this matters.
\end{assumption}

A continuation may designate a receiving verifier without designating the
future executor: the verifier is an enforcement component, whereas
\(x_{i+1}\) may remain unknown until the continuation is presented for
admission.

The single-consumption clause is stronger than the core PIC transition
semantics. Definition~\ref{def:valid-transition} does not prescribe one
successor per checkpoint. The present witness construction adds the property
that a predecessor checkpoint has at most one accepted successor.

This is a transition-admission property, not a message-delivery semantic.
Duplicate or at-least-once delivery is compatible with the same
guarantee of at most one accepted successor when repeated presentations of the
same predecessor are deduplicated by the atomic consumption decision.
Conversely, a realization may deliberately admit several successors, as in
fan-out; such a
realization does not provide the witness construction's property that the
checkpoint has at most one accepted successor.

In every case, each accepted continuation must independently satisfy the four
PIC transition obligations.

Single consumption does not require the predecessor to know the concrete
successor in advance. In the N+1 case, \(x_{i+1}\) may not yet exist, be
selected, or be provisioned when \(s_i\) emits its continuation. The uniqueness
decision is therefore made at admission: a valid candidate continuation is
evaluated by the receiving boundary and, in a single-consumption realization,
acceptance atomically consumes the predecessor. A realization that instead
required \(s_i\) to pre-bind the continuation to the future successor's
identity, workload key, holder identity, or communication channel would not
cover the unknown-successor case considered by the model.

\begin{proposition}[Relationship without successor-side possession]
\label{prop:witness}
Under Assumption~\ref{ass:witness}, and except with probability negligible in
\(\eta\), \(\PoR_\kappa(s_i,s_{i+1})\) implies
\(\PoR(s_i,s_{i+1})\). Moreover, if \(\kappa_{i+1}\) verifies, then \(\kappa_i\)
has at most one accepted successor, \(q_J(x_{i+1})\) is the profile of the
executor that realized \(s_{i+1}\), and the conditions \(C_{i+1}\subseteq C_i\),
\(\rho_i(\tau_i)\) and \(\rho_{i+1}\Rightarrow\rho_i\) hold. Verifying
\(\kappa_{i+1}\) involves no predicate \(\operatorname{Possess}(x_{i+1},k)\) for
any artifact \(k\).
\end{proposition}

\begin{proof}
Suppose \(\PoR_\kappa(s_i,s_{i+1})\). By unforgeability, except with
negligible probability, \(B_{i+1}\) issued \(\kappa_{i+1}\). Its payload's first
component is \(\operatorname{cid}(\kappa_i)\) for a verified \(\kappa_i\). A
verified checkpoint on the same payload but with different signature bytes has
the same identifier, and is the same checkpoint for every purpose of the
protocol. A verified checkpoint \(\kappa_i'\) on a different payload or from a
different boundary with \(\operatorname{cid}(\kappa_i')=\operatorname{cid}(\kappa_i)\)
would, since \(\operatorname{enc}\) is injective, be a collision in \(H\): any
execution of the system that produced one would yield an efficient collision
finder, so this happens with negligible probability. Except with
negligible probability, the predecessor binding therefore cannot be redirected
to a different valid checkpoint, and \(\kappa_{i+1}\) is bound to the values in
\(\kappa_i\). The two step-commitment equalities in the definition of
\(\PoR_\kappa\), together with collision resistance of \(h\), bind those
checkpoints to the named arguments \(s_i,s_{i+1}\) except with negligible
probability. For a chain rooted at \(\kappa_0\), root validity supplies the
corresponding base bindings and sound origin payload. By induction on \(i\), the
predecessor values in \(\kappa_i\) are correct. By protocol compliance,
\(B_{i+1}\) performed checks 1--5 before
issuing. Check~3 together with handoff observation gives
\(\operatorname{ObservedHandoff}(s_i,s_{i+1})\), and handoff adequacy gives
\(\PoR(s_i,s_{i+1})\). Check~2 together with
single-consumption/anti-replay gives that \(\kappa_i\) has at most one accepted
successor. Check~4 together with
non-PoP profile observation gives that \(q_J(x_{i+1})\) is correct. Check~5
gives \(C_{i+1}\subseteq C_i\) directly and \(\rho_{i+1}\Rightarrow\rho_i\) by
soundness of the syntactic criterion. For \(\rho_i(\tau_i)\), relational
evidence soundness and the lifting rules of Section~\ref{sec:relpred} turn
\(E_{i+1}\vdash\rho_i(\tau_i)\) into \(\rho_i(\tau_i)\). All three
are checked against predecessor values bound in \(\kappa_i\) and successor
values bound in \(\kappa_{i+1}\). Verification
reads \(\kappa_i\), \(\kappa_{i+1}\), the boundaries' public keys, and the
boundary-local or infrastructure evidence permitted by
Assumption~\ref{ass:witness}; nothing in it is held or presented by
\(x_{i+1}\).
\end{proof}

The substantive content of the proposition lies in handoff observation and
adequacy. The
construction converts an observation made by a compliant boundary into
verifiable evidence; it does not create causal knowledge that the boundary did
not have. Protocol compliance makes explicit what the previous paper leaves
implicit when it places compromise of the validator out of
scope~\cite{gallo2026poc}.

\paragraph{Realizing the assumptions without successor-side possession.}
The proposition shows that verification reads nothing from the successor. That
is not enough, because possession could re-enter through the assumptions
themselves. In practice, handoff observation and executor identification are
often realized with a key held by the workload, such as a SPIFFE SVID or an
attestation key presented by the executor. That is a proof of possession by the
successor. If the assumptions were realized that way, the claimed
no-successor-PoP realization would be circular. This is why handoff observation
is required to hold by the
boundary's own means, and why profile determination is required to be non-PoP
profile observation. Two architectural patterns illustrate how this requirement
may be met.
\begin{itemize}
    \item \emph{Launch and measure.} The boundary is the runtime that
    instantiates the executor. It launches \(x_{i+1}\) from an image it has
    measured, and assigns its profile at launch from that measurement and from
    policy. Identity and profile come from the launch record kept by the
    boundary, not from a key held by the workload.
    \item \emph{Co-located observation.} A runtime co-located with the
    executor, such as a hypervisor, a sandbox supervisor, or a proxy through
    which all input reaches the executor, delivers the continuation into the
    executor and observes its consumption directly. It knows which executor it
    hosts because it hosts it.
\end{itemize}
In both, the executor presents nothing: the boundary knows what it observes
because it controls the execution environment. Deployments that instead rely on
workload-held keys use successor-side PoP. The model still applies to them, but
the no-successor-PoP conclusion of
Theorem~\ref{thm:possession-independence} applies only when the Witness
Assumptions are in fact discharged by the deployment.

Possession does not disappear from the system: the boundaries control signing
keys. In this construction it moves from the successor to the witnesses, from
the presenter of authority to the root of trust that attests the causal
handoff. The successor is witnessed; it presents nothing. Verification of a
witness signature establishes the source and integrity of the attestation under
the trusted key; it does not by itself make the attested handoff or profile
true. Those conclusions use handoff adequacy and profile soundness. The
authorization rule then evaluates PoR, \(C_i\), \(\rho_i\), and the established
profile facts.

The same boundary also signs the profile facts \(q_J(x_{i+1})\), in the same
checkpoint and under the same key. Relational authorization therefore requires
no additional checkpoint-signing authority, provided the boundary that PoR
already trusts is also trusted to determine the profile facts. It does not
follow that no further trust is needed. Signing a statement does not make it
true, and the underlying profile evidence, such as the registry of measured
images, the hardware root of the launching platform, or the organizational
authority that assigns classes, may depend on additional trust roots. The
non-PoP profile observation assumption absorbs them.

\begin{theorem}[Conditional realization without successor-side possession]
\label{thm:possession-independence}
Under Assumption~\ref{ass:witness}, and except with probability negligible in
\(\eta\), the boundary-witnessed checkpoint construction realizes the premises
of Proposition~\ref{prop:semantic-possession-independence} for every transition
it admits, without requiring any successor executor to prove possession of an
artifact. Its admitted transitions therefore form a PIC-compatible realization
in the sense of Definition~\ref{def:pic-compatible}. Boundary signing keys
provide checkpoint evidence; external profile evidence may rely on additional
trust roots. This conclusion is relative to the Witness Assumptions, not an
unconditional proof that every required environmental observation can be
realized in every deployment.
\end{theorem}

\begin{proof}
Proposition~\ref{prop:witness} gives, under Assumption~\ref{ass:witness}, a
construction in which \(\PoR_\kappa\) soundly witnesses \(\PoR\), the successor
profile is sound, and \(C_{i+1}\subseteq C_i\), \(\rho_i(\tau_i)\), and
\(\rho_{i+1}\Rightarrow\rho_i\) hold. Relational evidence soundness lifts the
verified atomic facts to the admitted positive predicate. These are exactly the
requirements of Definition~\ref{def:pic-compatible}. Handoff observation by
the boundary's own means and non-PoP profile observation are hypotheses, so no
successor proves possession of anything within the construction. The resulting
admission rule is possession-independent by
Proposition~\ref{prop:semantic-possession-independence}.
\end{proof}

The launch-and-measure and co-located patterns above are architectural examples,
not a formal witness model proving the joint satisfiability of
Assumption~\ref{ass:witness}. Accordingly, the checkpoint argument is a proof
sketch in the computational model whose soundness is conditional on the seven
listed assumptions.

Theorem~\ref{thm:possession-independence} concerns admission of a transition by
\(\ValidR\). It yields a statement about an occurring resource effect only when
complete mediation connects effects to accepted steps; claims about effects
caused by a request additionally use causal attribution, exactly as in
Theorem~\ref{thm:causal-bound}.

\subsection{What the result does and does not claim}

Proposition~\ref{prop:semantic-possession-independence} records the semantic
non-necessity of PoP in the abstract rule. Theorem~\ref{thm:possession-independence}
is the separate, conditional realization result. Neither claims that PoP has no
security role.

PoP may still be useful for:
\begin{itemize}
    \item proving control of a private key;
    \item holder binding;
    \item authenticating an evidence producer;
    \item proving control of a continuation artifact;
    \item protecting transport or replay-sensitive mechanisms; and
    \item implementing a particular PoR construction.
\end{itemize}

The distinction is semantic:
\[
\boxed{
\begin{gathered}
\text{PoP may provide holder-binding}\\
\text{or authentication evidence;}
\end{gathered}
}
\]
whereas
\[
\boxed{
\begin{array}{c}
\text{PoR establishes causal continuation;}\\
\text{\(C_i\) bounds authority and \(\rho_i\) bounds}\\
\text{relational continuation.}
\end{array}
}
\]

Within this model, possession presented by the successor is therefore an
optional evidence or authentication-layer mechanism rather than a necessary
premise of the authorization rule. The conditional witness construction places
cryptographic possession at the boundary rather than at the successor; other
deployments may use PoP at either layer without making it a semantic premise.

\begin{remark}[Mechanism independence]\label{rem:mechanism}
Outside the witness construction of Section~\ref{sec:witness}, which is one
realization among many, no result in this paper refers to cryptography.
\(\PoR\), profile membership and the monotone state enter only as predicates;
signatures, hashes, attestation and proofs of possession are ways of producing
evidence for them. Proposition~\ref{prop:mechanism-invariance} proves invariance
under replacement of a profile-evidence mechanism only when the replacement is
sound and supplies the same profile. It does not prove that arbitrary evidence
mechanisms are interchangeable, nor does it remove the separate soundness
obligation for evidence of \(\PoR\) or relational atoms. PoR and PoC are modeled
as properties of the execution; PoP is one possible property established by an
evidence protocol.
\end{remark}

\section{Authorization Models over an Occurrence Domain}
\label{sec:observation}

The distinction between possession and relation can be stated without ordering
protocols by name. A possession-centered rule may be abstracted as
\[
\operatorname{Possess}(x,a)
\land
\operatorname{Grants}(a,o,r)
\Longrightarrow
\operatorname{Authorized}(x,o,r),
\]
whereas the PIC exercise rule evaluates a requested privilege against the
occurrence's authority context and, for a non-origin step, the relationally
valid transition that admitted it. That transition uses
\(\PoR(s_i,s_{i+1})\) and the monotone evolution
\(\Gamma_{i+1}\preceq\Gamma_i\). To compare these rules on one domain, let
\(\mathcal E\) be the space of admission-relevant evidence and verifier
context, and define
\[
\Omega
:=
\left\{
(\widehat{\ell}_i,x_i,o_i,r_i,\eta_i)
\ \middle|\
\begin{array}{l}
\widehat{\ell}_i\in\Lineages,\quad x_i=\exec(s_i),\\
(o_i,r_i)\in O\times R,\quad \eta_i\in\mathcal E
\end{array}
\right\}.
\]
An element of \(\Omega\) is a complete authorization occurrence for comparison
purposes: its history prefix contains the endpoint state \((C_i,\rho_i)\), and
the remaining coordinates record its executor, requested privilege, and the
evidence, trusted observations and contextual facts available to the admission
boundary. Thus \(\eta_i\) may include presented artifacts, verifier-local or
infrastructure observations, externally retrieved facts, revocation state,
authenticated-channel facts, policy and environment context, trusted time, or
boundary-local state. It records inputs to admission, not semantic truth:
\(\PoR\) and the other semantic premises remain distinct from evidence used to
establish them. This auxiliary domain does not alter
Definition~\ref{def:valid-transition}; it makes explicit which coordinates a
decision procedure retains or forgets.

\begin{definition}[Observation-based authorization model]
An authorization model \(M\) supplies an observation map, a policy on the
observation, and a transition-admissibility relation. The observation map is a
surjection
\[
\operatorname{Obs}_M:\Omega\longrightarrow Y_M,
\]
where, without loss of generality, \(Y_M=\operatorname{Obs}_M(\Omega)\). Its
kernel equivalence is
\[
\omega\sim_M\omega'
\quad\Longleftrightarrow\quad
\operatorname{Obs}_M(\omega)=\operatorname{Obs}_M(\omega').
\]
The policy is a map \(\bar P_M:Y_M\to\{0,1\}\), inducing the occurrence-level
admission verdict
\[
P_M=\bar P_M\circ\operatorname{Obs}_M
\colon\Omega\longrightarrow\{0,1\}.
\]
Finally,
\[
\operatorname{Trans}_M\subseteq\Omega\times\Omega
\]
is the state-propagation or transition-admissibility relation. The Boolean
verdict \(P_M\) and the binary relation \(\operatorname{Trans}_M\) are distinct:
the former decides a requested privilege at one occurrence, while the latter
constrains evolution between occurrence states.
\end{definition}

Thus an observation does not describe everything a protocol message can encode.
It describes the distinctions on \(\Omega\) that a particular admission rule
actually preserves.

\begin{definition}[Observation refinement]
Write
\[
M\preceq_{\mathrm{obs}}N
\]
when there is a map \(f:Y_N\to Y_M\) such that
\[
\operatorname{Obs}_M=f\circ\operatorname{Obs}_N.
\]
Then \(M\) is observationally no finer than \(N\): every distinction retained
by \(M\) is available to \(N\), although \(N\) need not use it in its policy.
\end{definition}

\begin{proposition}[Observation factorization and refinement]
\label{prop:observation-refinement}
Let \(M\) and \(N\) be observation models on \(\Omega\). Then:
\begin{enumerate}
    \item if \(P_M=\bar P_M\circ\operatorname{Obs}_M\), then
    \[
    \omega\sim_M\omega'\Longrightarrow P_M(\omega)=P_M(\omega');
    \]
    \item
    \[
    M\preceq_{\mathrm{obs}}N
    \quad\Longleftrightarrow\quad
    \ker(\operatorname{Obs}_N)
    \subseteq
    \ker(\operatorname{Obs}_M);
    \]
    \item if \(M\preceq_{\mathrm{obs}}N\), every policy that factors through
    \(\operatorname{Obs}_M\) also factors through
    \(\operatorname{Obs}_N\).
\end{enumerate}
\end{proposition}

\begin{proof}
The first statement is Proposition~\ref{prop:factorization} applied to the
kernel quotient of \(\operatorname{Obs}_M\). If
\(\operatorname{Obs}_M=f\circ\operatorname{Obs}_N\), equality under
\(\operatorname{Obs}_N\) implies equality under \(\operatorname{Obs}_M\),
which proves the forward kernel inclusion. Conversely, suppose the kernel
inclusion holds. Define
\[
f(\operatorname{Obs}_N(\omega)):=\operatorname{Obs}_M(\omega).
\]
The inclusion makes \(f\) well-defined, and surjectivity onto the chosen images
gives \(\operatorname{Obs}_M=f\circ\operatorname{Obs}_N\). Finally,
\(P_M=\bar P_M\circ f\circ\operatorname{Obs}_N\), proving the third statement.
\end{proof}

For PIC, the requested-privilege coordinate is consumed by the occurrence-level
exercise verdict
\[
P_{\mathrm{PIC}}(\omega_i)=1
\quad\Longleftrightarrow\quad
s_i\text{ validly exercises }(o_i,r_i)
\]
in the sense of Definition~\ref{def:exercise}. Equivalently,
\((o_i,r_i)\in C_i\) and either \(i=0\), or the transition entering \(s_i\) is
relationally valid. The origin tuple is already an admitted root occurrence;
selection of \(C_0\) by the bootstrap policy remains outside scope.
\(P_{\mathrm{PIC}}\) is the occurrence-level exercise verdict; validity of the
complete admitted prefix is represented separately by the sequence of
transitions in \(\operatorname{Trans}_{\mathrm{PIC}}\). This
Boolean verdict is not \(\operatorname{Trans}_{\mathrm{PIC}}\): the latter,
defined below, governs propagation from one authorization state to the next.

\begin{proposition}[Observation sufficiency for exact realization]
\label{prop:observation-sufficiency}
Let \(T:\Omega\to\{0,1\}\) be a target semantic authorization verdict. There
exists \(\bar T:Y_M\to\{0,1\}\) such that
\[
T=\bar T\circ\operatorname{Obs}_M
\]
if and only if \(T\) is invariant on every equivalence class induced by
\(\operatorname{Obs}_M\). Equivalently,
\[
\ker(\operatorname{Obs}_M)\subseteq\ker(T),
\]
where
\[
\ker(T):=\{(\omega,\omega'):T(\omega)=T(\omega')\}.
\]
Consequently, if
\(\operatorname{Obs}_M(\omega)=\operatorname{Obs}_M(\omega')\) but
\(T(\omega)\neq T(\omega')\), no policy factoring solely through
\(\operatorname{Obs}_M\) realizes \(T\) exactly on both occurrences.
\end{proposition}

\begin{proof}
The kernel inclusion is exactly invariance of \(T\) on the fibers of
\(\operatorname{Obs}_M\). Since \(Y_M=\operatorname{Obs}_M(\Omega)\), define
\[
\bar T(\operatorname{Obs}_M(\omega)):=T(\omega).
\]
Invariance makes \(\bar T\) well-defined and gives the required factorization;
the converse follows immediately from the factorization. If the displayed
indistinguishable pair has different target verdicts, every factored policy
assigns one verdict to both and hence disagrees with \(T\) on at least one.
This is Proposition~\ref{prop:factorization} applied to the kernel quotient of
\(\operatorname{Obs}_M\), not an additional independent assumption.
\end{proof}

The order \(\preceq_{\mathrm{obs}}\) compares information resolution only. It
does not say that \(N\) is more secure than \(M\), that its evidence is sounder,
or that its admitted transitions satisfy stronger invariants.
It is a preorder on observation maps. Identifying maps whose kernels are equal
makes the induced relation a partial order on observation-kernel classes.

Because \(\eta_i\) may itself expose causal history, dropping only
\(\widehat{\ell}_i\) would not in general forget lineage. Fix a
history-erasure observation
\[
\epsilon_{\neg L}:\mathcal E\longrightarrow\mathcal E_{\neg L}
\]
that, for the comparison being made, removes or abstracts away evidence whose
semantic purpose is to distinguish causal histories. Such evidence includes
lineage or predecessor identifiers, checkpoint hashes, continuation receipts,
PoR evidence, handoff evidence, and equivalent causal-history facts. There need
not be one canonical erasure for every evidence representation; the projection
is parameterized by the chosen \(\epsilon_{\neg L}\).

Define the corresponding lineage-forgetting projection by
\[
\begin{aligned}
\mathcal Z_{\neg L}:={}&
X\times O\times R\times\mathcal E_{\neg L}\\
&{}\times\mathcal P(O\times R)
\times(\Pred(\Transitions_J)/{\equiv_{\mathrm{sem}}}),
\end{aligned}
\]
and
\[
\begin{aligned}
\pi_{\neg L}:\Omega&\longrightarrow\mathcal Z_{\neg L},\\
(\widehat{\ell}_i,x_i,o_i,r_i,\eta_i)
&\longmapsto
(x_i,o_i,r_i,\epsilon_{\neg L}(\eta_i),
C_i,[\rho_i]_{\mathrm{sem}}).
\end{aligned}
\]

\begin{corollary}[Lineage-forgetting limitation]
\label{cor:lineage-forgetting}
If \(\operatorname{Obs}_M=g\circ\pi_{\neg L}\), then any policy factoring
through \(\operatorname{Obs}_M\) gives the same verdict to occurrences that
agree on the retained coordinates but differ in their explicit causal-history
prefixes and in evidence erased by \(\epsilon_{\neg L}\). Consequently,
separating occurrences solely by those causal-history differences requires an
observation that does not factor through \(\pi_{\neg L}\).
\end{corollary}

\begin{proof}
Equal values under \(\pi_{\neg L}\) imply equal values under
\(\operatorname{Obs}_M\); the first part of
Proposition~\ref{prop:observation-refinement} then gives equal policy verdicts.
The final statement is the contrapositive.
\end{proof}

This corollary restates, on the common domain \(\Omega\), the quotient argument
of Remark~\ref{rem:instance}. The explicit history coordinate and the evidence
components designated by \(\epsilon_{\neg L}\) as history-revealing are erased
by construction. It does not say that any named protocol family is
necessarily lineage-forgetting: a realization can retain sound causal-history
facts in its observation.

For a narrower artifact-local comparison, fix an evidence projection
\[
\epsilon_{\mathrm{art}}:\mathcal E\longrightarrow\mathcal E_{\mathrm{art}}
\]
that retains exactly the artifact-local or possession-local evidence selected
for that comparison, while excluding any lineage-local or current-authority
information carried in \(\eta_i\). The projection below also omits the history
prefix and its endpoint authorization state. There need not be one canonical
choice for all protocols. Define
\[
\begin{aligned}
\pi_{\mathrm{art}}:\Omega
&\longrightarrow X\times O\times R\times\mathcal E_{\mathrm{art}},\\
(\widehat{\ell}_i,x_i,o_i,r_i,\eta_i)
&\longmapsto(x_i,o_i,r_i,\epsilon_{\mathrm{art}}(\eta_i)).
\end{aligned}
\]

\begin{corollary}[Artifact-local indistinguishability]
\label{cor:artifact-local}
Suppose \(\operatorname{Obs}_M=g\circ\pi_{\mathrm{art}}\). If there are
\(\omega,\omega'\in\Omega\) such that
\[
\pi_{\mathrm{art}}(\omega)=\pi_{\mathrm{art}}(\omega'),
\qquad
P_{\mathrm{PIC}}(\omega)=1,
\qquad
P_{\mathrm{PIC}}(\omega')=0,
\]
then no policy factoring through \(\operatorname{Obs}_M\) implements
\(P_{\mathrm{PIC}}\) exactly on both occurrences. Such a policy either admits
both, producing a false admission relative to \(P_{\mathrm{PIC}}\), or rejects
both, producing a false rejection relative to \(P_{\mathrm{PIC}}\).
\end{corollary}

\begin{proof}
The factorization through \(\pi_{\mathrm{art}}\) makes the two observations
equal. Proposition~\ref{prop:observation-sufficiency} then shows that every
policy based solely on that observation gives them the same verdict, whereas
the target gives different verdicts. The two possible common Boolean verdicts
give the stated alternatives.
\end{proof}

An important witness pair has the same executor-visible artifact observation
and requested privilege, but endpoint authority contexts \(C\) and \(C'\) with
\((o,r)\in C\) and \((o,r)\notin C'\), while the remaining premises make the
first exercise valid. The corollary is an observational insufficiency result
about exact realization, not a claim that every artifact system permits mixing: a
reject-all policy avoids false admission but incurs the stated false rejection.
Nor is it a limitation of artifact carriers as such. A capability, token, or
assertion that soundly exposes the missing lineage-local authority or causal
facts lies in a finer observation class, and the corollary does not apply to
that pair.

An authorization model includes \(\bar P_M\) as well as its observation and
transition relation. Section~\ref{sec:observation} compares it on two axes by
the signature
\[
\Sigma(M):=
(\ker(\operatorname{Obs}_M),\operatorname{Trans}_M).
\]
The first coordinate records observational resolution; the second records
state-propagation invariants. This signature is deliberately not the whole
authorization behavior: two models can have the same \(\Sigma(M)\) but
different \(\bar P_M\), and hence different occurrence-level verdicts.
Observation refinement and transition admissibility remain separate.

For
\[
\begin{aligned}
\omega_i&=(\widehat{\ell}_i,x_i,o_i,r_i,\eta_i),\\
\omega_{i+1}&=(\widehat{\ell}_{i+1},x_{i+1},o_{i+1},r_{i+1},\eta_{i+1}),
\end{aligned}
\]
write
\[
\widehat{\ell}_i\prec_L^{\,1}\widehat{\ell}_{i+1}
\quad\Longleftrightarrow\quad
\widehat{\ell}_{i+1}
=\widehat{\ell}_i\mathbin{\|}
\langle(s_{i+1},C_{i+1},\rho_{i+1})\rangle,
\]
where \(\mathbin{\|}\) denotes sequence concatenation. Define the lifted PIC
relation by
\[
\begin{aligned}
&(\omega_i,\omega_{i+1})
\in\operatorname{Trans}_{\mathrm{PIC}}\\
&\quad\Longleftrightarrow\quad
\widehat{\ell}_i\prec_L^{\,1}\widehat{\ell}_{i+1}\\
&\qquad\land\;
\ValidR(s_i,C_i,\rho_i,
s_{i+1},C_{i+1},\rho_{i+1}).
\end{aligned}
\]
The immediate-prefix condition requires the two elements of \(\Omega\) to be
consecutive endpoints of the same authorization-aware history. The second
condition is exactly the four-conjunct rule of
Definition~\ref{def:valid-transition}:
\[
\PoR(s_i,s_{i+1}),\qquad
C_{i+1}\subseteq C_i,\qquad
\rho_i(\tau_i),\qquad
\rho_{i+1}\Rightarrow\rho_i.
\]
Immediate-prefix compatibility is a well-formedness condition of the relation
after it has been lifted to \(\Omega\times\Omega\); it is not a fifth PIC
authorization conjunct and does not modify
Definition~\ref{def:valid-transition}.
The first axis asks which occurrences an admission rule can distinguish. The
second asks which state changes it accepts. A model may be fine on one axis and
weak or incomparable on the other; no total ranking follows.

The authorization question can now be stated precisely: not only what
authority an actor possesses, but whether the observed occurrence is a valid
continuation under an authorization state that has not expanded. Identity,
credentials, capabilities and PoP may contribute coordinates or evidence to
\(\operatorname{Obs}_M\). They do not by their carrier type alone determine
either the kernel of that observation or \(\operatorname{Trans}_M\).

\section{A Realization of the Premises: PIC-X}
\label{sec:picx}

The results of this paper are stated over abstract predicates and do not depend
on any implementation. This section records how one open-source implementation,
PIC-X~\cite{picx}, part of the PIC protocol~\cite{picprotocol}, discharges the
restricted subset of premises identified below. Its purpose is an engineering
illustration, not a proof or a complete witness of the model. The description was checked on
25 September 2026 against the PIC-X source at release \texttt{v0.2.3}
(commit \texttt{b93b890}), which builds on the \texttt{pic-protocol} crate, version
\texttt{0.2.3}. The PIC-X release tag is an implementation snapshot, not an
assumption of any theorem in this paper.

PIC-X takes the layered route of Proposition~\ref{prop:layers}, not the witness
route of Section~\ref{sec:witness}. The successor authenticates with
possession, and each checkpoint is signed by the realm that settles the
transition. Possession-based signatures have two evidence-producing roles:
they support workload/profile authentication through the attester binding, and
they demonstrate control of the bytes used to advance a continuation artifact.
The predicate \(\operatorname{Possess}\) itself is not a primitive of the
abstract relational authorization decision.

\paragraph{A difference in terminology.}
PIC-X uses the name ``Proof of Relationship'' for the evidence that
authenticates the workload: an SD-JWT~\cite{sdjwt} issued by an attester that
the realm trusts, which binds the workload's public key through its
confirmation claim. In the terms of this paper, that evidence belongs to the
authentication layer. The causal relationship that this paper calls \(\PoR\) is
only partially evidenced in PIC-X by the transition, which references the exact
predecessor checkpoint; this does not establish an observed handoff.
PIC-X therefore describes this implementation level as \emph{artifact-linked}:
the label records the strength of the available evidence and is not a new
semantic definition of \(\PoR\).

\paragraph{Settlement.}
The workload submits a candidate transition and signs it, together with the
candidate continuity and the candidate token, with the key that the attester
bound. The realm then verifies, in order: the attester's presentation and the
workload's three signatures; that the predecessor checkpoint verifies under the
realm's own key; the hash of the exact predecessor bytes, the position
progression, the challenge chain, the lineage identifier and the expiry; the
conformance of the disclosed claims to the execution contract; non-expansion;
the configured revocation hook, which is currently a no-op; and local policy.
Only then does it issue the next checkpoint, signed by the realm.

\begin{table*}[t]
\centering
\small
\begin{tabular}{@{}p{0.27\linewidth}p{0.67\linewidth}@{}}
\toprule
Model element & PIC-X \texttt{v0.2.3} \\
\midrule
Profile \(q_J(x)\) &
Claims disclosed in the attester-issued SD-JWT presentation, verified at
settlement and recorded with the attester's identity for attribution. \\
Relation \(\sim_j\) &
Not a separate object. Equality of the disclosed value of claim \(j\), from a
trusted attester, is the induced equivalence: the kernel of the claim function. \\
Authentication of membership &
Attester-issued SD-JWT binding the workload key in its confirmation claim; the
workload signs the transition, the candidate continuity and the candidate token
with that key. Key-binding JWTs are not used, and are rejected. \\
\(\PoR(s_i,s_{i+1})\) in this paper &
Only partially realized. The transition carries the SHA-256 hash of the exact
predecessor checkpoint bytes, position \(n+1\), challenge continuity and the
checkpoint's lineage identifier, and the predecessor checkpoint must verify
under the realm's own key. This evidences possession of the predecessor
checkpoint, not an observed handoff: handoff observation and adequacy are not
realized. \\
\(C_i\) and \(C_{i+1}\subseteq C_i\) &
Context of authority: identity context and invariants, attenuated by removal
bitmaps and checked by subset inclusion. \\
\(\rho_i\) and \(\rho_{i+1}\Rightarrow\rho_i\) &
Execution contract: key--value and membership constraints on the successor's
disclosed claims. The predecessor's contract governs admission; additions are
conjunctive only, and no entry can be removed or weakened. Unary and
conjunctive; binary atoms and disjunction are not implemented. \\
Checkpoint signing &
Each checkpoint is a COSE object signed by the realm; accepting a predecessor is
a signature check under the realm's published keys. \\
Freshness &
SD-JWT validity claims, with bounded clock skew. The absolute lineage expiry is
fixed in the initial checkpoint and preserved across advancement. A checkpoint
is advanceable only before that expiry and while the realm key that signed it
is published. The protocol crate provides a revocation hook at settlement, but
PIC-X \texttt{v0.2.3} configures none. No credential status list in this
version. \\
Linearity and replay &
Stateless: a checkpoint is not retired when a successor is issued, so fan-out
is supported. There is no single-consumption register. \\
\bottomrule
\end{tabular}
\caption{Correspondence between the model and PIC-X \texttt{v0.2.3}, with
\texttt{pic-protocol} \texttt{0.2.3}.}
\label{tab:picx}
\end{table*}

\paragraph{The execution contract as a unary envelope.}
The execution contract is a restricted envelope. Its atoms depend only on the
successor's profile, \(\rho_i(u,v)=\bigwedge_k R_k(v)\), and they are combined by
conjunction only. Addition-only accumulation is exactly the syntactic refinement
\(\rho_{i+1}=\rho_i\land\sigma\) of Section~\ref{sec:refinement}. The contract
is evaluated from the predecessor checkpoint, which matches the design choice
that \(\rho_i\) governs hop \(i\). Atoms that relate predecessor and successor,
such as the separation-of-duty atom of Section~\ref{sec:sod}, are beyond this
version.

\paragraph{Why positive predicates suit selective disclosure.}
The evidence monotonicity of the positive language has a direct reading here. A
holder that withholds a disclosure removes evidence, and removing evidence
cannot produce an admission that full disclosure would deny. Selective
disclosure therefore cannot be used to widen admission. PIC-X behaves this way:
a contract entry that is not disclosed is a rejection. The same property
explains why constraints must be positive attested claims: an SD-JWT cannot
establish the absence of a claim by withholding it.

\paragraph{Partial profiles.}
Selective disclosure makes the verified profile partial: the boundary learns
only the components that were disclosed, while \(q_J(x)\) in the model is total.
The gap is the one the evidence relation \(\vdash\) addresses. The boundary
admits on \(E\vdash\rho_i(\tau_i)\), where \(E\) contains the verified
disclosures. Soundness of \(E\) gives \(\rho_i(\tau_i)\) for the true profile,
and undisclosed components can only prevent admission.

\paragraph{Where the realization departs from the model.}
The following departures affect no theorem, but they bound what PIC-X
\texttt{v0.2.3} realizes.
\begin{itemize}
    \item \emph{Successor-side possession.} PIC-X authenticates the successor
    with a key it holds. This is the layered route of
    Proposition~\ref{prop:layers}: possession enters two evidence layers,
    authentication and continuation evidence, but not the decision rule. The
    conditional no-successor-PoP
    realization of Theorem~\ref{thm:possession-independence} is the construction
    of Section~\ref{sec:witness}, not PIC-X.
    \item \emph{Fan-out.} PIC-X admits sibling branches from one checkpoint,
    while this paper models linear chains. Every root-to-leaf path is a linear
    chain, and every result of this paper applies along it. No branch can import
    authority from another, because each successor is attenuated from its own
    predecessor. The single-consumption/anti-replay assumption of the witness
    construction does not hold in PIC-X, and is needed only for the claim that
    a checkpoint has at most one accepted successor, which PIC-X deliberately
    does not provide.
    \item \emph{Relationship as possession of the checkpoint.} PIC-X does not
    observe a handoff. The relationship is evidenced by the workload's signature
    over a transition that references the exact predecessor bytes and continues
    its challenge. The checkpoint travels inside the settled token, and its
    challenge is in clear. Any workload that is attested by a trusted attester,
    conforms to the execution contract and obtains the settled token can
    therefore advance the lineage, whether or not the token was handed to it.
    The damage is bounded by the checkpoint's authority and by the contract,
    but the relationship reduces to possession of a continuation artifact. The
    present stateless construction does not close this gap. In the models
    considered here, closing it requires trusted handoff observation, additional
    state, or an equivalent mechanism; the construction of
    Section~\ref{sec:witness} leaves that observation to a separate component.
    \item \emph{No request binding.} The protocol crate offers a
    request-binding hook, but PIC-X's artifact-linked policy configures it to
    accept without establishing request binding because no trusted request or
    transport occurrence is available at that boundary. A checkpoint is
    therefore not tied to the request it continues.
    \item \emph{Unconstrained settled token.} The settled token carries no
    audience binding and no successor-key binding in this version. The
    unknown-successor setting prevents pre-binding the continuation to a
    concrete future executor identity or key when that executor is not yet
    known, but it does not preclude binding to a receiving verifier, service,
    trust domain, audience, or other receiver class known in advance. The
    absence of such audience binding is therefore a design property of this
    realization, not a consequence of the N+1 model. As configured, the
    settled continuation is bearer-like within the class of attested,
    contract-conforming workloads accepted by the realm.
    \item \emph{No revocation.} No revocation check is configured; a checkpoint
    is bounded only by the lineage expiry and by the retention of the realm key.
    \item \emph{Attribution.} Complete mediation and causal attribution, the
    premises of Theorem~\ref{thm:causal-bound} and
    Corollary~\ref{thm:cd}, are properties of the deployment around PIC-X,
    such as resource servers that accept only settled tokens, not of PIC-X
    alone. Given the first item, PIC-X does not by itself supply causal
    attribution, including for dependencies carried through persistent state,
    asynchronous work, covert channels or side channels.
\end{itemize}

\paragraph{Planned implementation boundary as of 25 September 2026.}
These omissions are separated deliberately from the guarantees of
\texttt{v0.2.3}. The planned next implementation phase introduces trusted
Execution Contexts at the consumption boundary. Because a message publisher
may not know whether zero, one or many executors will consume a message, and a
successor key may not yet exist, the design does not pre-bind a checkpoint to a
future successor identity or key. Once a consumer is selected, an Execution
Context can observe the concrete handoff and issue a signed receipt, or
equivalent evidence, committing to the predecessor or candidate digest, a
fresh occurrence identifier, the consumed request or message, and the locally
authenticated executor profile. Authenticated channel facts may be included
where the deployment can soundly observe them.

Such evidence is intended to strengthen artifact linkage into observed-handoff
evidence, provide request binding, and support causal attribution for explicit
mediated handoffs. Complete mediation additionally requires the protected
resource and execution boundaries to reject effects not admitted through that
path; PIC-X alone cannot establish it for hidden state, covert channels, or
side channels. Optional replay protection, if introduced, must be scoped to an
observed occurrence or delivery rather than globally retiring the predecessor,
so that intentional fan-out remains possible. Revocation remains a separate
mechanism, with explicit state and availability semantics. Per-claim attester
trust, rather than the current issuer-wide trust, is independent hardening.
None of these planned mechanisms is assumed to exist in \texttt{v0.2.3}, and
none is used by the paper's theorems.

\paragraph{What the realization does not change.}
No theorem of this paper depends on PIC-X. PIC-X realizes authority
attenuation, a unary conjunctive envelope, and profile soundness resting on the
configured attesters (issuer-wide in this version); the linear results apply
per path. It does
not realize handoff observation and adequacy, the witness's
single-consumption/anti-replay property, request binding, or
revocation, binary relational atoms, disjunction, complete mediation, or
side-channel attribution, so it does not discharge the Witness Assumptions of
Section~\ref{sec:witness} nor, by itself, the premises of
Theorem~\ref{thm:causal-bound}.

\section{Related Work}
\label{sec:related}

Section~\ref{sec:observation} gives the comparison criterion: protocol names do
not define semantic classes. For the two-axis comparison, a concrete
realization is positioned by the kernel of the observation actually used for
admission and by its admitted transition relation; this signature is not its
complete Boolean authorization policy. No protocol family below is asserted to
be a set-theoretic subset of PIC. The rows summarize mechanism profiles; they do not assign one canonical
observation map to every named protocol or prove every pairwise kernel
inclusion. The same carrier format may occur in different rows of
Table~\ref{tab:families}, and the rows are not a security ranking.

\begin{table*}[t]
\centering
\caption{Authorization families on the two comparison axes. Entries describe
the stated profile, not every use of the named standards or artifacts.}
\label{tab:families}
\small
\begin{tabular}{@{}>{\raggedright\arraybackslash}p{0.16\textwidth}
>{\raggedright\arraybackslash}p{0.23\textwidth}
>{\raggedright\arraybackslash}p{0.14\textwidth}
>{\raggedright\arraybackslash}p{0.16\textwidth}
>{\raggedright\arraybackslash}p{0.16\textwidth}@{}}
\toprule
Family & Observation retained at admission & Holder binding & Compositional attenuation & Causal occurrence continuity \\
\midrule
Artifact-, token-, or assertion-centered
& Artifact or assertion, claims, requested privilege and selected verifier context
& Not intrinsic to bearer profiles
& Profile-dependent
& Not intrinsic when causal history is unobserved \\
\addlinespace
Sender-constrained or holder-bound
& Artifact/claim observation plus holder- or key-binding evidence; some profiles bind a proof to request properties
& Required by the stated holder-bound profile
& Profile-dependent
& Not implied by holder/key binding or local request binding \\
\addlinespace
Attenuating capability or delegation construction
& Designated capability/reference or authority-bearing artifact, together with restrictions and ancestry/state exposed by the construction
& Construction-dependent
& Enforced when descendants can only preserve or restrict authority
& Construction-dependent; authority or delegation ancestry does not by itself imply semantic PoR \\
\addlinespace
PIC-compatible authorization
& Requested privilege, current authorization state, and sound occurrence, causal-history and executor-profile facts used by the admission rule
& Optional evidence mechanism
& \(C_{i+1}\subseteq C_i\) and \(\rho_{i+1}\Rightarrow\rho_i\)
& \(\PoR(s_i,s_{i+1})\) is required \\
\bottomrule
\end{tabular}
\end{table*}

\paragraph{Artifact-, token-, and assertion-centered profiles.}
This class includes OAuth bearer-token profiles~\cite{rfc6749,rfc6750}, JWT
access artifacts when used as bearer credentials~\cite{rfc7519}, and SAML
assertions used with bearer subject confirmation~\cite{samlcore}. The class is
defined narrowly by an admission rule whose observation is artifact-local, not
by all uses of OAuth, JWT or SAML. If its actual observation factors through
\(\pi_{\mathrm{art}}\), Corollary~\ref{cor:artifact-local} applies to any
PIC-valid/PIC-invalid pair that the projection identifies; if it factors
through \(\pi_{\neg L}\), Corollary~\ref{cor:lineage-forgetting} gives the
corresponding history-erasure result. The standards remain free to carry or
reference additional evidence; a profile that soundly observes causal
occurrence or current lineage authority belongs to a finer or different class.

\paragraph{Sender-constrained and holder-bound profiles.}
DPoP is sender-constrained and binds each proof to HTTP request properties as
specified by its profile~\cite{rfc9449}. SAML Holder-of-Key
assertions~\cite{samlhok} and mutual-TLS certificate-bound
credentials~\cite{rfc8705} establish holder or key binding without thereby
establishing causal occurrence continuity. None of holder binding, key binding,
or local request binding alone implies \(\PoR(s_i,s_{i+1})\). Holder binding
therefore induces, relative to the corresponding bearer-style observation, a
finer observation without by itself introducing the causal-predecessor
coordinate. A concrete profile may nevertheless add sound continuity evidence
and, if it enforces the required transition obligations, become a
PIC-compatible realization.

\paragraph{Attenuating capability and delegation constructions.}
Capabilities unite designation and authority and address the classical
confused-deputy problem when invocation uses the designated authority rather
than ambient authority~\cite{dennis1966programming,miller2003capability}.
Macaroons append contextual caveats~\cite{birgisson2014macaroons}; Biscuit and
UCAN provide attenuating delegation structures~\cite{biscuitspec,ucanspec};
recent agent-delegation proposals likewise combine predecessor commitments
with attenuated grants~\cite{asor2026delegation,atakora2026sadp}. These facts
establish compositional restriction in their respective constructions; they do
not place every construction in one observation class. Authority attenuation
does not by itself establish causal continuity, and artifact or delegation
ancestry is not automatically semantic occurrence continuity. A construction
that soundly establishes equivalent causal continuation and enforces all four
obligations of Definition~\ref{def:pic-compatible} is a capability-based
PIC-compatible realization. PIC does not replace its carrier or policy
language.

\paragraph{PIC-compatible authorization.}
This is a semantic class, not a token format. Membership requires soundly
established \(\PoR\) and profile facts, together with enforcement of
\(C_{i+1}\subseteq C_i\), \(\rho_i(\tau_i)\), and
\(\rho_{i+1}\Rightarrow\rho_i\) at admission. Merely encoding the relevant
fields is insufficient, while any mechanism family can qualify when its
admission semantics discharge those obligations. Complete Mediation and
Causal Attribution remain separate premises for conclusions about effects, as
Definition~\ref{def:pic-compatible} states. The witness construction's
single-consumption rule is likewise a property of that realization, not a
fifth PIC transition conjunct.

\paragraph{Adjacent formalisms.}
ABAC classifies subjects, objects, operations and environments for policy
evaluation~\cite{hu2014abac}; the maps \(q_J\) provide one such classification
of executors, while PIC additionally propagates a refining relational envelope.
Dynamic and history-based separation of duty constrain roles or prior
participants~\cite{ferraiolo2001rbac,simon1997sod}, and history-based access
control makes execution history a policy input~\cite{abadi2003history}.
Proof-carrying and distributed authorization can prove premises used by an
admission rule~\cite{appel1999pca,bauer2005distributed}. These formalisms are
complementary: each can supply observations or checks, while the comparison
signature \(\Sigma(M)\) records which distinctions and transition invariants
the resulting system actually uses.

The quotient construction remains agnostic to the meaning of each executor
equivalence. Trust-domain, attestation, workload and organizational classes
belong in relations on \(X\); occurrence-specific continuation belongs in
\(\PoR\) or the carried history. What matters is not whether a carrier is
called a token, assertion, capability or proof, but which distinctions its
admission rule preserves and which state transitions it accepts.

The semantic questions can therefore be separated. A bearer-style admission
rule asks which artifact and claims are presented and accepted. A
sender-constrained or PoP mechanism additionally asks who demonstrates control
of associated key material and, in profiles such as DPoP, binds that proof to
request properties. An attenuating capability or delegation construction may
also govern how designated or carried authority is restricted as it is derived.
PIC additionally asks whether this concrete occurrence validly continues this
predecessor occurrence under authority and relational state that have not
expanded.

Capabilities determine what authority is carried and may be attenuated.
PoP systems determine who can demonstrate control of authority-bearing material.
PIC additionally determines whether this particular execution occurrence is a valid causal continuation in which that authority remains admissible.

Here ``who'' denotes the holder or key controller identified relative to the
profile's trusted binding; it is not a claim that PoP alone establishes a
person's identity, and holder identity is not causal occurrence identity.
These dimensions are not mutually exclusive and do not form a total security
order. A capability may use PoP; a token may carry attenuating authority; either
may carry lineage-sensitive evidence. Authority attenuation is not causal
continuity, and artifact ancestry is not automatically semantic occurrence
continuity. The distinction is in the
semantic domain: PIC makes causal occurrence an explicit authorization
dimension and composes it with authority and relational non-expansion. A token,
assertion, capability, caveat, or logic-based construction that soundly
establishes the causal-continuation and profile facts and enforces the four
obligations of Definition~\ref{def:pic-compatible} is PIC-compatible
irrespective of its carrier.

\section{Discussion}

\subsection{Why quotient spaces matter}

The quotient construction is not merely notation. It separates concrete
identity from authorization-relevant identity.

Suppose \(x\neq y\) but
\[
x\sim_j y.
\]
Then any policy intentionally defined only over
\[
\Rel_j=X/{\sim_j}
\]
must treat \(x\) and \(y\) equally. If the policy must distinguish them, then
\(\sim_j\) was too coarse and a finer relational view must be introduced.

Multiple quotient spaces therefore provide a disciplined way to state which
executor differences matter to authorization and which do not.

\subsection{Refinement instead of replacement}

A lineage need not carry one fixed relational predicate forever. It may refine
the predicate:
\[
R_1\lor R_2
\;\succeq_R\;
R_1
\;\succeq_R\;
R_1\land R_3.
\]

This is the relational analogue of authority attenuation:
\[
\{\text{read},\text{write}\}
\supseteq
\{\text{read}\}.
\]

The shared principle is non-expansion.

\subsection{Dynamic separation of duty}
\label{sec:sod}

The relational envelope constrains which executors may act along a lineage, not
only which authority they carry. It therefore expresses constraints between
executors that authority attenuation alone cannot state.

Let \(a,j\in J\). Let \(\sim_a\) classify executors by role, and let
\(A\subseteq\Rel_a\) be the approver classes. For the relation \(\sim_j\) and
profiles \((u,v)\in\Transitions_J\),
with \(u_j\) the \(j\)-th component of \(u\), define the atom
\[
\mathrm{SoD}_j(u,v)
:\Longleftrightarrow
v_a\notin A
\;\lor\;
u_j\neq v_j.
\]
The atom is a fixed subset of \(\Transitions_J\) and is evaluated directly from
the verified profile components.
It holds trivially when the successor is not an approver; when it is, the
approver must lie in a different \(\sim_j\)-class from its predecessor.

A lineage whose proposal step \(s_i\) is followed by an approval step can refine
its envelope at the proposal step to
\[
\rho_i:=\rho_{i-1}\land\mathrm{SoD}_j.
\]
This is a monotone refinement, checked by the receiving boundary at acceptance.
Because every later envelope implies \(\rho_i\), no descendant can drop the
constraint. The result is dynamic separation of duty, the four-eyes principle,
expressed as a refinement of the relational envelope.

Authority attenuation cannot state this constraint: proposer and approver may
carry the same authority context, and the constraint concerns who exercises it.
An admission rule observing only a bearer artifact cannot evaluate the
constraint unless the necessary subject relation is also soundly exposed to
that rule. Mechanisms with such an observation can express it, as classified in
Section~\ref{sec:related}. Here it is one refinement of \(\rho\).

The atom relates adjacent executors, and the adjacent case lies within the
formalism. The non-adjacent case does not. A predicate sees only
\(\tau_i=(q_J(x_i),q_J(x_{i+1}))\), and a proposer class recorded in a
checkpoint is a state variable outside that domain. A formal treatment extends
the state to
\[
\Gamma_i=(C_i,\rho_i,m_i),
\]
where \(m_i\) is a write-once memory attested by the boundary, such as the
proposer's class, and lets predicates range over \(M\times\Transitions_J\). The
closure arguments use only extension inclusion and transitivity, so they carry
over to the extended domain unchanged. We leave that development to future
work.

\subsection{The role of causality}

Neither authority attenuation nor relational refinement alone establishes
continuity. The two sequences could be monotone but belong to unrelated
executions.

PoR is what binds the monotone states to one causal history:
\[
s_0\rightarrow s_1\rightarrow\cdots\rightarrow s_n.
\]

Thus the model has three inseparable components:
\begin{enumerate}
    \item causal linkage by PoR;
    \item monotone privilege authority; and
    \item monotone relational admissibility.
\end{enumerate}

\subsection{The lineage axis}

Previous work represented an authority occurrence as
\[
(o,r,\ell)\in O\times R\times\mathcal L_{\mathrm{prev}},
\]
where \(\ell\) identifies the causal lineage. The present model does not replace
that axis.

Instead, it refines the structure carried \emph{along} a lineage. The lineage
prefix
\[
\widehat{\ell}_i
\]
grows, while the attached state
\[
\Gamma(\widehat{\ell}_i)=(C_i,[\rho_i]_{\mathrm{sem}})
\]
shrinks in the product attenuation order.

At the modeling level, this may be read as an ontological shift in what bears
authorization state: the relevant object is not merely an executor or artifact,
but an execution occurrence embedded in causal history. This is an
interpretation, not an additional theorem; its mathematical content is the
history-indexed state \(\Gamma(\widehat{\ell}_i)\) and the transition rule of
Definition~\ref{def:valid-transition}.

It is therefore mathematically more precise to say:
\begin{quote}
continuity is monotone in the authorization state carried along the lineage,
not that the lineage itself is decreasing.
\end{quote}

\section{Limitations and Future Work}
\label{sec:limitations}

The model is intentionally minimal and several questions remain open.

First, the equivalence relations \(\sim_j\) are treated as given. A practical
system must define who is authorized to assert them, how evidence is verified,
how classes change over time, and how revocation affects class membership.

Second, the paper assumes a common relational vocabulary \(J\). Heterogeneous
systems may use different executor classifications at different boundaries. A
general treatment would require sound translations between quotient spaces,
analogous to authority translation between heterogeneous operation/resource
spaces.

Third, the formal histories and proofs are linear. The core transition rule,
however, does not prescribe single consumption: its obligations can be checked
independently on each root-to-leaf path of a fan-out. The present witness adds
at most one accepted successor per checkpoint and therefore excludes fan-out.
Repeated delivery or presentation is compatible with this property when the
atomic consumption decision deduplicates attempts. A fan-out realization
instead deliberately admits several successors; each branch must independently
satisfy the valid-transition obligations, but the realization does not provide
the witness construction's property that the checkpoint has at most one
accepted successor. A unified treatment of forks, joins, fan-in and DAG-shaped
executions still requires a partial-order generalization of both authority
attenuation and relational refinement.

The same linearity excludes multiple-cause effects across distinct histories.
Because \(\operatorname{Lineage}\) is functional, Causal Attribution cannot
hold simultaneously for
\(\operatorname{CausedBy}(e,req_1)\) and
\(\operatorname{CausedBy}(e,req_2)\) when
\(\widehat{\ell}_{req_1}\neq\widehat{\ell}_{req_2}\). Jointly caused effects
require both a DAG or fan-in causal semantics and an explicit composition rule
for the participating origin contexts and relational envelopes. Intersection
of origin contexts is one possible conservative design, not a result of the
present model; no composition rule is chosen here.

Fourth, PoR remains largely abstract. The paper proves properties conditional
on sound verification of the causal relationship. The boundary-witnessed
construction of Section~\ref{sec:witness} gives a no-successor-PoP realization
relative to its Witness Assumptions, but does not independently prove their
joint satisfiability. It is only a computational proof sketch and is not
prescribed as the only construction.

Fifth, the exclusion of the confused deputy rests on complete mediation and
causal attribution (Assumptions~\ref{ass:mediation}
and~\ref{ass:attribution}). The model does not decide whether an effect was
caused by a request. The boundary-witnessed construction realizes attribution
for explicit mediated handoffs only; causality that flows through mutable
memory, queues, databases, storage, caches, shared files, timers, asynchronous
callbacks or background workers requires a stronger execution model,
information-flow tracking for the modeled channels~\cite{myers1997decentralized},
or a further assumption.
Covert channels and side channels are not resolved by the witness construction:
they may influence control or data flow without appearing as a mediated
handoff, and ordinary information-flow tracking need not capture all of them.
The confused-deputy result remains valid conditional on Causal Attribution, but
the construction does not establish that assumption for arbitrary hidden
channels. An error in attribution is a failure of the trusted computing base,
not a violation of the continuity invariant.

The witness construction also separates assumptions that deployments must
realize independently: handoff observation, non-PoP profile observation,
whose evidence may rest on further trust roots, and
single-consumption/anti-replay.

A deployment that requires at most one accepted successor for a checkpoint
needs an atomic consumption decision or equivalent coordination. When the
consumption register is shared across boundaries, a globally unique acceptance
decision requires consistent coordination; under a network partition,
preserving that safety property may require delaying or rejecting admission
rather than allowing two boundaries to accept the same predecessor
independently. This is a realization-level distributed
consistency/availability trade-off, not a theorem of the abstract PIC
semantics.

Delivery semantics are separate from transition-admission semantics. An
at-least-once transport may present the same continuation multiple times while
an atomic consumption register deduplicates those attempts and still yields at
most one accepted successor. Conversely, a realization may deliberately
support fan-out and admit multiple valid successors. Such a realization retains
the PIC obligations on each admitted branch but does not provide the witness
construction's property that the checkpoint has at most one accepted
successor.

The N+1 unknown-executor property is orthogonal to this choice. The concrete
successor need not be known when the predecessor acts; a single-consumption
realization is compatible with that model provided uniqueness is enforced at
admission without requiring pre-identification or pre-binding of the future
executor. A continuation may designate a verifier without designating that
future executor.

Dynamic separation of duty between non-adjacent steps requires the state
extension sketched in Section~\ref{sec:sod}.

Sixth, Proposition~\ref{prop:semantic-possession-independence} is the immediate
semantic result: PoP is not a premise of the abstract authorization rule once
the required relationship facts are given. Theorem~\ref{thm:possession-independence}
is instead a conditional construction showing how those premises can be
realized without successor-side PoP under the Witness Assumptions. A deployment
may still choose PoP for engineering reasons, including authentication, key
confirmation, replay resistance, or evidence transport.

The present model orders authorization state along an already materializing
causal execution. A fuller temporal treatment must also account for executors,
facts, and outcomes that do not yet exist when an earlier authorization
decision is made. The N+1 unknown-executor problem captures the case in which
the concrete successor materializes only after the predecessor has acted,
while the N+1 unknown-outcome problem captures the stronger case in which an
authorization-relevant fact or concrete outcome itself becomes available only
during execution~\cite{picNPlus1Executor,picNPlus1Outcome}. A future temporal
authorization space may therefore need to represent admissible future
occurrences or outcomes before their concrete witnesses materialize. We leave
that extension to future work.

Finally, the model does not claim that every authorization policy should be
relational. It identifies a relational authorization structure for discrete
causal executions where authority continuity matters.

\section{Conclusion}

Proof-of-Continuity established that authority can be propagated through a
causal execution by combining Proof of Relationship with monotone authority
attenuation. This paper develops the relational structure implicit in that
model.

Executors are classified through authorization-relevant equivalence relations:
\[
\sim_j\subseteq X\times X.
\]
Each relation induces a quotient:
\[
\Rel_j=X/{\sim_j},
\]
which identifies physically different executors whenever their differences are
irrelevant to the selected authorization characteristic.

Multiple relational quotients coexist, and their combined profile
\[
q_J(x)
\]
provides the relational coordinates of an executor. Predicates over transitions
between such profiles compose with conjunction and disjunction, and implication
defines an attenuation order:
\[
\rho_{i+1}\Rightarrow\rho_i.
\]

A continuous execution therefore carries two monotone authorization dimensions:
\[
C_{i+1}\subseteq C_i
\]
and
\[
\rho_{i+1}\Rightarrow\rho_i.
\]

At the same time, causal history accumulates:
\[
\widehat{\ell}_i\preceq_L\widehat{\ell}_{i+1}.
\]

The resulting structure is:
\[
\boxed{
\begin{gathered}
\text{history grows while authority}\\
\text{and relational freedom shrink.}
\end{gathered}
}
\]

Dual monotonicity gives combined closure. No valid descendant may introduce a
privilege absent from the origin authority context, and no valid hop may
introduce a relational transition excluded by the origin relational envelope.
Under two explicit assumptions, complete mediation and causal attribution, the
request-bounded-effect theorem excludes any accepted request-caused effect
outside the origin authority context, and the confused-deputy result follows as
a corollary. This does not preclude a buggy deputy from attempting such an
operation; it precludes acceptance of the resulting effect under the stated
premises. The enforcement boundary construction covers attribution only for
explicit handoffs under its Witness Assumptions, and not for arbitrary state,
covert or side channels. The authorization model is at the same time
strengthened with relational closure.

PIC therefore separates authority possession from authority admissibility. A
deputy may hold authority from unrelated sources, but that authority cannot
become a valid exercise of the current lineage merely because the deputy
executes one of its steps. Confused-deputy authority mixing is excluded by
construction within a valid lineage; global exclusion of request-caused effects
additionally depends on complete mediation and causal attribution.
This is a semantic claim, not an exclusivity claim: other authorization
mechanisms can obtain the same no-mixing result by adopting an admission rule
that preserves the same lineage-local authority and relational constraints.
Definition~\ref{def:pic-compatible} makes this criterion explicit.

Most importantly, Proposition~\ref{prop:semantic-possession-independence}
records that the logical authorization rule contains no necessary
successor-side possession premise. If PoR and relational-profile facts are
soundly verifiable, PoP may supply holder-binding evidence or be used by one
particular witness mechanism, but it is not required as the semantic basis of
transition admission. Theorem~\ref{thm:possession-independence} separately
shows that the checkpoint construction supports a no-successor-PoP realization
only relative to its explicit Witness Assumptions. Claims about resource effects
additionally require complete mediation and causal attribution.

The distinction can be stated compactly:
\[
\boxed{
\begin{gathered}
\text{possession can evidence control;}\\
\text{PoR establishes causal continuity.}
\end{gathered}
}
\]
Equivalently, in terms of the two layers:
\[
\boxed{
\begin{array}{c}
\text{sound evidence establishes}\\
\text{executor profiles;}\\
\text{PoR establishes causal continuation;}\\
\text{authorization combines PoR,}\\
\text{the profiles, \(C_i\), and \(\rho_i\) to decide}\\
\text{what the occurrence may exercise}\\
\text{in that history.}
\end{array}
}
\]

The relevant authorization subject is therefore an execution occurrence in
causal history, not a holder or artifact alone. The common-domain analysis of
Section~\ref{sec:observation} makes the claim independent of carrier names. An
observation map determines which occurrences a policy can distinguish; the
Boolean policy determines the requested privileges it admits; and a separate
transition relation determines which state evolutions it accepts. The
observation-sufficiency result characterizes exactly when an observation can
realize a target verdict and when an indistinguishable valid/invalid pair makes
exact realization impossible. Separately, PIC transition admission enforces
PoR, authority attenuation, relational satisfaction and relational refinement.
Observation refinement is only an information order and gives no total security
ranking among bearer, sender-constrained, capability and PIC-compatible
families. It provides a criterion for comparing concrete observations; it does
not assert canonical observation maps or pairwise kernel inclusions for every
named protocol family.

Capabilities, credentials, caveats, assertions and PoP may all supply evidence
or carry authority. Any such construction is PIC-compatible precisely when its
admission semantics soundly establish the required continuation and profile
facts and enforce the four transition obligations of
Definition~\ref{def:pic-compatible}. Evidence of PoR is not the relationship
itself, just as profile evidence is not the authorization decision. The carrier
therefore does not determine the semantic class; the preserved distinctions and
transition invariants do. Capability attenuation, holder or key binding, and
artifact ancestry may each contribute to such a realization, but none by
itself determines semantic causal continuation of a particular occurrence.

PIC changes the domain in which this evidence is consumed: authorization
applies to execution occurrences situated in causal histories.
Consequently,
\[
\boxed{
\begin{array}{c}
\text{possessing authority}\\
\neq\\
\text{being the occurrence permitted to exercise it}\\
\text{in this causal history}
\end{array}
}.
\]

\section*{Acknowledgment of Prior Model}

This paper extends the Proof-of-Continuity and Provenance Identity Continuity
model developed in earlier work~\cite{gallo2026poc}. The new
contribution is the executor quotient construction, the relational predicate
order, dual monotonicity, relational closure, the reading of the earlier
projection result as a quotient instance, the occurrence-domain observation
and refinement framework, and the possession-independence result together with
a conditional construction that satisfies its premises without successor-side
possession when the Witness Assumptions hold.

\appendix
\section{Worked Example}
\label{app:example}

Let \(J=\{\mathrm{org},\mathrm{role},\mathrm{prin}\}\), where
\(\sim_{\mathrm{prin}}\) classifies executors by the principal on whose behalf
they act. Several executor instances may share a principal, so its classes are
not singletons. Profiles are triples
\(u=(u_{\mathrm{org}},u_{\mathrm{role}},u_{\mathrm{prin}})\), and the approver
classes are \(A=\{\mathrm{approver}\}\subseteq\Rel_{\mathrm{role}}\). Define the
atom
\[
\mathrm{SameOrg}(u,v):\Longleftrightarrow u_{\mathrm{org}}=v_{\mathrm{org}},
\]
and take \(\mathrm{SoD}_{\mathrm{prin}}\) from Section~\ref{sec:sod}. Both atoms
are decidable from the profiles.

\paragraph{The lineage.}
A request enters at \(s_0\), realized by \(x_0\) with profile
\((\mathrm{acme},\mathrm{requester},\mathrm{alice})\), with
\[
\begin{aligned}
C_0={}&\{(\mathrm{read},r),(\mathrm{write},r),\\
&(\mathrm{approve},p)\},\\
\rho_0={}&\mathrm{SameOrg}.
\end{aligned}
\]
Step \(s_1\) is realized by \(x_1\) with profile
\((\mathrm{acme},\mathrm{proposer},\mathrm{bob})\). The hop from \(s_0\) is
admitted, since \(\rho_0\) holds on \((q_J(x_0),q_J(x_1))\). At \(s_1\), bob
proposes a change and refines the envelope:
\[
C_1=C_0,
\qquad
\rho_1=\mathrm{SameOrg}\land\mathrm{SoD}_{\mathrm{prin}}.
\]
The refinement \(\rho_1\Rightarrow\rho_0\) holds syntactically, by conjunction.
Step \(s_2\) is an approval by executor \(x_2\). It presents evidence, for
instance an SD-JWT from an issuer trusted for organization, role and principal,
disclosing \(\mathrm{org}=\mathrm{acme}\), \(\mathrm{role}=\mathrm{approver}\)
and \(\mathrm{prin}=\mathrm{carol}\). The boundary verifies it and fixes
\(q_J(x_2)=(\mathrm{acme},\mathrm{approver},\mathrm{carol})\).

\paragraph{Acceptance.}
With \(\tau_1=(q_J(x_1),q_J(x_2))\), the boundary checks \(\rho_1(\tau_1)\). The
organizations agree. The successor is an approver, and its principal differs from
the predecessor's, since \(\mathrm{bob}\neq\mathrm{carol}\). Both conjuncts
hold. The boundary also checks \(C_2=\{(\mathrm{approve},p)\}\subseteq C_1\)
and \(\rho_2=\rho_1\), a trivial refinement. The hop is accepted. Conditional
on the same fixed occurrence-level PoR and authorization state, any sound
profile-evidence mechanism that establishes the same successor profile induces
the same executor-dependent relational verdict.
Proposition~\ref{prop:mechanism-invariance} does not assert that replacing the
executor or occurrence preserves PoR or creates another admitted execution.

\paragraph{Rejections.}
Four variants fail, each for a different reason.
\begin{enumerate}
    \item Another executor instance acting for bob approves:
    \(\tau=((\mathrm{acme},\mathrm{proposer},\mathrm{bob}),
    (\mathrm{acme},\mathrm{approver},\mathrm{bob}))\). The successor is an
    approver and the principals coincide, so \(\mathrm{SoD}_{\mathrm{prin}}\)
    fails. Separation is between principals, not instances: a second process of
    the same principal is not a second pair of eyes.
    \item An approver from another organization: \(\mathrm{SameOrg}\) fails.
    \item An approver that withholds the organization disclosure: the evidence
    does not support \(\mathrm{SameOrg}\), so \(E\nvdash\rho_1(\tau_1)\) and the
    hop is rejected. By evidence monotonicity, no withholding can turn a
    rejection into an admission.
    \item An approver whose transition carries
    \((\mathrm{delete},r)\notin C_1\): non-expansion fails.
\end{enumerate}
In none of these cases does holding a valid credential change the outcome: the
decision is taken over the profile, the authority context and the lineage.

In PIC-X \texttt{v0.2.3}, the fixed-value part of this policy can be written as
an execution-contract entry such as \(\mathrm{org}=\mathrm{acme}\). The
relational atoms \(\mathrm{SameOrg}\) and \(\mathrm{SoD}_{\mathrm{prin}}\),
which relate predecessor and successor, need binary atoms, which that version
does not implement.

\end{document}